\documentclass[aps, prd, superscriptaddress, 
nofootinbib, showpacs, preprintnumbers]{revtex4}
\usepackage{graphicx}
\usepackage{amsmath}
\usepackage{bm}
\usepackage{multirow}
\def\fsl#1{\setbox0=\hbox{$#1$}                 
   \dimen0=\wd0                                 
   \setbox1=\hbox{/} \dimen1=\wd1               
   \ifdim\dimen0>\dimen1                        
      \rlap{\hbox to \dimen0{\hfil/\hfil}}      
      #1                                        
   \else                                        
      \rlap{\hbox to \dimen1{\hfil$#1$\hfil}}   
      /                                         
   \fi}                                         %
\newcommand{\tr}{\mbox{tr}}
\newcommand{\Tr}{\mbox{Tr}}

\newcommand{\diag}{\mbox{diag}}

\newcommand{\VEV}[1]{\langle #1 \rangle}

\begin{document}
\title{Gluonic phases, vector condensates, and exotic hadrons in dense QCD}
\author{E.V. Gorbar}
  \email{gorbar@bitp.kiev.ua}
  \affiliation{
       Bogolyubov Institute for Theoretical Physics,
       03143, Kiev, Ukraine}
\author{Michio Hashimoto}
  \email{michioh@eken.phys.nagoya-u.ac.jp}
\affiliation{Department of Physics, Nagoya University, Nagoya,
464-8602, Japan}
\author{V.A. Miransky}
  \email{vmiransk@uwo.ca}
   \altaffiliation[On leave from ]{
       Bogolyubov Institute for Theoretical Physics,
       03143, Kiev, Ukraine}
\affiliation{
Department of Applied Mathematics, University of Western
Ontario, London, Ontario N6A 5B7, Canada\\
and Yukawa Institute for Theoretical Physics, Kyoto University,
Kyoto 606-8502, Japan}
\date{\today}
\preprint{UWO-TH-07/03}

\begin{abstract}
We study the dynamics in phases with vector condensates of gluons
(gluonic phases) in dense two-flavor quark matter. These phases 
yield an example of dynamics in which the Higgs mechanism is
provided by condensates of gauge (or gauge plus scalar) fields.
Because vacuum expectation values of spatial components of vector fields break
the rotational symmetry, it is natural to have a spontaneous
breakdown both of external and internal symmetries in this
case. In particular, by using the Ginzburg-Landau approach, we
establish the existence of a gluonic phase with
both the rotational symmetry and the electromagnetic
$U(1)$ being spontaneously broken. In other words, this phase 
describes an anisotropic medium in which the color and electric
superconductivities coexist. 
It is shown that this phase
corresponds to a minimum of the Ginzburg-Landau potential
and, unlike the two-flavor superconducting (2SC) phase,
it does not suffer from the chromomagnetic instability.
The dual (confinement) description of its dynamics is
developed and it is shown that there are light exotic vector hadrons
in the spectrum, some of which condense.
Because most of the initial symmetries
in this system are spontaneously broken, its dynamics is very rich.
\end{abstract} 

\maketitle

\section{Introduction}

It is expected that at sufficiently high baryon density, cold quark
matter should be in a color superconducting state~
\cite{Barrois:1977xd,CSC,CSC1}.
On the other hand, it was suggested long ago that quark matter might
exist inside the central region of compact stars~\cite{quark_star}.
This is one of the main reasons why the dynamics of
the color superconductivity 
has been intensively studied (for reviews, see Ref.~\cite{review}).

Bulk matter in compact stars should be in $\beta$-equilibrium,
providing by weak interactions, 
and be electrically and color neutral.
The electric and color neutrality conditions play a crucial role
in the dynamics of quark pairing
\cite{Iida:2000ha,Alford:2002kj,Steiner:2002gx,Huang:2002zd,
Huang:2003xd,Abuki:2004zk,Ruster:2005jc}. Also, in the dense quark matter,
the strange quark mass cannot be neglected.
These factors lead
to a mismatch $\delta\mu$ between the Fermi momenta of
the pairing quarks.

As was revealed in Ref. \cite{Huang:2004bg},
the 
gapped (2SC) and gapless (g2SC)
two-flavor color superconducting phases suffer from a chromomagnetic
instability connected with the presence of imaginary 
Meissner masses of gluons.
While the 8th gluon has an imaginary Meissner mass only
in the g2SC phase, with the diquark gap
$\bar{\Delta} < \delta\mu$ (an intermediate coupling regime),
the chromomagnetic instability for the 4-7th gluons appears also in
a strong coupling regime, with
$\delta\mu < \bar{\Delta} < \sqrt{2}\delta\mu$.
Later a chromomagnetic instability was also found in
the three-flavor gapless color-flavor locked (gCFL)
phase~\cite{Casalbuoni:2004tb,Alford:2005qw,Fukushima:2005cm}.

Meissner and Debye masses are screening (and not pole) ones.
It has been recently revealed in Ref. \cite{Gorbar:2006up}
that the chromomagnetic instabilities 
in the 4-7th and 8th gluonic channels correspond to two very
different tachyonic spectra of plasmons.
It is noticeable that while
(unlike the Meissner mass)
the (screening) Debye mass for an electric mode remains real for
all values of $\delta\mu$ both in the 2SC and g2SC phases 
\cite{Huang:2004bg}, 
the tachyonic plasmons occur both for the magnetic and electric modes
\cite{Gorbar:2006up}. The
latter is important since it clearly shows that this instability is connected
with vectorlike excitations: Recall that
two magnetic modes correspond to two 
transverse components of a plasmon, and one electric mode
corresponds to its longitudinal component. 
This form of the plasmon spectrum leads to the unequivocal conclusion
about the existence of vector condensates of gluons
in the ground state 
of two flavor quark matter with 
$\bar{\Delta} < \sqrt{2}\delta\mu$, thus
supporting the scenario with gluon condensates (gluonic phase)
proposed in letter \cite{Gorbar:2005rx}.
While the analysis in Ref.~\cite{Gorbar:2005rx} was done only in
the vicinity of the critical point 
$\delta\mu \simeq \bar{\Delta}/\sqrt{2}$,
a numerical analysis of the gluonic phase far away of the scaling
region ~was considered in Refs.
\cite{Fukushima:2006su,Kiriyama:2006ui,He:2006vr}.

At intermediate energy scales of the order of the diquark condensate
$\bar{\Delta} \sim {\cal O}(\mbox{50MeV})$, the analysis of QCD dynamics
is very hard.
Hence the phenomenological Nambu-Jona-Lasinio (NJL) model
plays a prominent role in the analysis in dense quark matter
(for recent extensive studies of dense QCD in this approach, see
Refs. \cite{Abuki:2004zk,Ruster:2005jc}).
The NJL model is usually regarded as a low-energy
effective theory in which massive gluons are integrated out.
The situation with dense quark matter is however quite different from
that in the vacuum QCD. We will introduce gluonic degrees of freedom
into the NJL model because in the 2SC/g2SC phase
the gluons of the unbroken $SU(2)_c$ 
subgroup of the color $SU(3)_c$
are left as massless, and, under certain conditions considered below,
some other gluons can be also very light. This yields 
the gauged NJL model. 

Also, because of the presence of matter, the running of
the QCD coupling constant dramatically changes.
As was shown in Ref.~\cite{Rischke:2000cn}, 
the confinement scale $\Lambda'_{\rm QCD}$ in the 2SC phase is
essentially smaller than the typical scale of
the diquark condensate, 
$\Lambda'_{\rm QCD} \lesssim {\cal O}(\mbox{10MeV})$ or
even much smaller.
This justifies introducing the gluonic degrees of freedom
into the NJL model at the energy scale around and below
$\bar{\Delta}$ (and larger than $\Lambda'_{\rm QCD}$),
as was done in Ref.~\cite{Gorbar:2005rx}. 
At such scales, the dynamics in the gluonic
phase corresponds to the Higgs picture.

As we will discuss in this paper, the description of the 
infrared dynamics in the gluonic phase depends on the relation
between two scales: $\Lambda'_{\rm QCD}$ and the value of
of the vector gluon condensates. If the latter is  larger
than $\Lambda'_{\rm QCD}$,
then the Higgs picture is appropriate even in the infrared
region, similarly as it
happens in the electroweak theory.
Indeed, 
when the $SU(2)_c$ gauge symmetry becomes completely broken
by the dynamics with a characteristic scale being
essentially larger than 
$\Lambda'_{\rm QCD}$,
the strong coupling dynamics presented
in the 2SC solution at the scale of order $\Lambda'_{\rm QCD}$ 
is washed out. However,
if this characteristic
scale is $\alt \Lambda'_{\rm QCD}$,
the confinement picture should be used. 
Although such a dynamics is not under control,
the structure of the global symmetry should be
the same both in the Higgs and confinement phases, if the Higgs
fields (both vector and scalar ones) are assigned to the
fundamental representation of the gauge group.
In particular, the global charges of hadrons in the confinement
picture should correspond to those of fundamental fields
in the Higgs picture.
This is one of the manifestations of the
complementarity
principle~\cite{Osterwalder:1977pc,Fradkin:1978dv,Dimopoulos:1980hn}.
Using this approach,
it will be shown the existence of exotic hadrons in the gluonic phase.
Moreover, we will see that 
dynamics with vector gluon condensates in the Higgs phase
correspond to dynamics with condensates of exotic vector mesons
in the confinement one. 

From the viewpoint of quantum field theory,
it is quite natural to expect the existence of vector condensates
of gluons
in order to remove such instabilities as the chromomagnetic one or
those connected with tachyonic plasmons.
In Ref. \cite{Gorbar:2005rx}, homogeneous, i.e., independent of spatial
coordinates, vector gluon condensates were considered. At the same time,
because the condensates of spatial components of gluon fields
break the rotational symmetry, such condensates are anisotropic.
On the other hand, inhomogeneous condensates 
for diquark fields were studied in 
Refs.
\cite{Alford:2000ze,Reddy:2004my,Giannakis:2004pf,
Hong:2005jv,Kryjevski:2005qq,Huang:2005pv}.
For example, the 
Larkin-Ovchinnikov-Fulde-Ferrell (LOFF) phase~\cite{LOFF}
in QCD~was considered in Refs. \cite{Alford:2000ze,Giannakis:2004pf}. 
In this phase
a diquark condensate is inhomogeneous. Some solutions with
inhomogeneous diquark condensates, including the single
plane-wave LOFF phase, 
can be considered as a special case of homogeneous vector
condensates of gluons with the field strength being zero 
\cite{Gorbar:2005rx,Gorbar:2005tx,Hashimoto:2006mn}.
However, when the latter is not zero, as in the case of
the gluonic phase \cite{Gorbar:2005rx}, homogeneous
vector gluon condensates
cannot be traded for inhomogeneous diquark ones. 
Moreover, as was suggested
in Ref. \cite{Gorbar:2006up}, inhomogeneous vector condensates 
of gluons could
exist in the g2SC and gCFL phases.

The paper is organized as follows. In Sec. \ref{sec2},
a renormalizable model with condensates of gauge fields
is considered. This essentially soluble model yields a proof that
such dynamics exist indeed. In Sec. \ref{sec3}, the gauged NJL
model is described.  Because the diquark gap
$\bar{\Delta}$
breaks the initial color
$SU(3)_c$ symmetry to the $SU(2)_c$ one, it is useful to decompose
the fields in the gauged NJL model with respect to the  $SU(2)_c$ subgroup.
This decomposition is considered in  Sec. \ref{sec4}. 
Section \ref{gluon} is the central in this paper. In its four
subsections, the dynamics in the gluonic phase 
is described in detail. One cannot exclude that besides this
phase, other phases with vector condensates of gluons may exist in the
gauged NJL model.
In Sec. \ref{general}, we classify possible sets of homogeneous
gluon condensates and typical symmetry breaking patterns
in the corresponding phases in this model. We also
describe the sets of operators
relevant for constructing the Ginzburg-Landau effective theories
for these dynamics. In particular,
as a consequence of this analysis, it is shown
that the ansatz for gluon condensates used for the gluonic phase
\cite{Gorbar:2005rx} is self-consistent. In Sec. \ref{conclusion},
the main results of the paper are summarized. In Appendices 
\ref{A} and \ref{B} , some useful formulas and relations are
derived.

\section{Renormalizable model for dynamics with vector condensates}
\label{sec2}

Since a dynamics with vector condensates is a rather new ``territory'',
it would be important to have an essentially soluble model which
would play the same role for such a dynamics
as the linear $\sigma$
models play for the conventional dynamics with spontaneous symmetry
breaking with condensates of scalar fields. Fortunately, such a model
exists: it is the gauged linear $SU(2)_L\times U(1)_Y$
$\sigma$-model (without fermions) with a chemical potential for
hypercharge $Y$ \cite{sigmamodel}. 
\footnote{Ungauged linear $SU(2)_L\times U(1)_{Y}$
$\sigma$-model with a chemical potential for
hypercharge \cite{MS}
is a toy model for the description of the dynamics of 
the kaon condensate in high density QCD \cite{BS}. 
In particular, it realizes the phenomenon with abnormal
number of Nambu-Goldstone (NG) bosons \cite{MS},
when spontaneous
breakdown of continuous symmetries leads to a lesser number of
NG bosons than that required by the Goldstone
theorem (for a recent discussion of this model, see
Ref. \cite{Brauner}).}

Let us briefly describe this model. It will be very useful 
for better understanding the dynamics in the gluonic phase. 
Its Lagrangian density reads (we use the metric
$g^{\mu\nu}= \mbox{diag}(1, -1, -1, -1))$:
\begin{equation} {\cal L}=-\frac{1}{4}F^{(a)}_{\mu\nu}F^{\mu\nu(a)}-
\frac{1}{4}F^{(Y)}_{\mu\nu}F^{\mu\nu(Y)} +
[(D_{\nu}-i\mu_Y\delta_{\nu0})\Phi]^{\dag}
(D^{\nu}-i\mu_Y\delta^{\nu0})\Phi-m^2\Phi^{\dag}
\Phi-\lambda(\Phi^{\dag}\Phi)^2,
\label{Lagrangian}
\end{equation}
where the covariant derivative $D_{\mu}=\partial_{\mu}-igA_{\mu}-
(ig^{\prime}/2)B_{\mu}$, $\Phi$ is a complex doublet field
$\Phi^T=(\varphi^+,\varphi_0)$, and the chemical
potential $\mu_Y$ is provided by external conditions (to be specific, we take
$\mu_Y >0$). Here
$A_{\mu}=A_{\mu}^{(a)}\tau^a/2$ are $SU(2)_L$ gauge fields ($\tau^a$ are 
three Pauli matrices) and the field strength
$F^{(a)}_{\mu\nu}=\partial_{\mu}A_{\nu}^{(a)}-
\partial_{\nu}A^{(a)}_{\mu} + g\epsilon^{abc}A^{(b)}_{\mu}A^{(c)}_{\nu}$.
$B_{\mu}$ is a $U_{Y}(1)$ 
gauge field with the field strength $F^{(Y)}_{\mu\nu}  =
\partial_{\mu}B_{\nu}-\partial_{\nu}B_{\mu}$. The hypercharge of the doublet
$\Phi$ equals +1. This model has the same
structure as the electroweak theory without fermions and with the chemical
potential for hypercharge $Y$. 
Note that the terms with the chemical potential
are $SU(2)_L \times U(1)_Y$ (and not $SU(2)_{L}\times SU(2)_R$)
symmetric. This follows from the fact that
the hypercharge generator $Y$ is $Y = 2I^{3}_{R}$ where $I^{3}_{R}$
is the
third component of the right handed isospin generator. 
Henceforth
we will omit the subscripts $L$ and $R$, allowing various
interpretations of the $SU(2)$.

The model is renormalizable and for small coupling 
constants $g$, $g^{\prime}$ and $\lambda$, the tree approximation 
is reliable there.
Because the chemical potential explicitly breaks
the Lorentz symmetry, the symmetry of the model is 
$SU(2) \times U(1)_Y \times SO(3)_{\rm rot}$.
As was shown in Ref. \cite{sigmamodel}, for sufficiently large
values of the chemical potential $\mu_Y$, the condensates of
both the scalar doublet $\Phi$ {\it and} the gauge field $A_\mu$ occur. 
The ground state solution is given by
\begin{equation}
|\VEV{W^{(-)}_z}|^2 = \frac{\mu_Y v_0}{\sqrt{2}g}-\frac{v_0^2}{4},
\quad \VEV{A^{(3)}_0} = \frac{v_0}{\sqrt{2}}, \quad 
\VEV{\Phi^T}=(0,v_0),
\label{vacuum}
\end{equation}
where
\begin{equation}
v_0=\frac{\sqrt{(g^2+64\lambda)\mu_{Y}^2-
8(8\lambda-g^2)m^2}-3g\mu_Y}{\sqrt{2}
(8\lambda-g^2)} \,,
\label{v0}
\end{equation}
$W^{(\mp)}_{\mu}=\frac{1}{\sqrt{2}}(A_{\mu}^{(1)} \pm iA_{\mu}^{(2)})$,
$\Phi^T=(\varphi^+,\varphi_0)$,
and the vacuum expectation values of 
all other fields are equal to zero \cite{footnote1}.   
It is clear that this solution implies that the initial symmetry
$SU(2) \times U(1)_Y \times SO(3)_{\rm rot}$ is spontaneously
broken down to $SO(2)_{\rm rot}$. In particular, the
electromagnetic $U(1)_{em}$, with electric charge 
$Q_{em} = I_3 + Y/2$,
is spontaneously broken by the condensate
of $W$ bosons, i.e., electric superconductivity takes place in
this medium.
\footnote{Note that because the $U(1)_Y$ symmetry
is local, for a nonzero chemical potential $\mu_Y$
one should introduce a source term $B_0J_0$ in Lagrangian density
(\ref{Lagrangian}) in order to make the system neutral with respect to
hypercharge $Y$. This is necessary since otherwise in such a system 
thermodynamic equilibrium could not be
established. The value of the background hypercharge density $J_0$
(representing very heavy particles) is determined from the requirement that
$B_0=0$ is a solution of the equation of motion  
for $B_0$ (the Gauss's law)
\cite{sigmamodel,Kapusta}. There exists an alternative description of
this dynamics in which a background hypercharge density $J_0$ is
considered as a free parameter and $\mu_Y$ is taken to be zero. 
Then Gauss's law will define the vacuum expectation value 
$\VEV{B_0}$. It is not difficult to check that these two approaches
are equivalent if the chemical potential $\mu_Y$ in the first 
approach is taken to be equal to the value
$\frac{g'}{2}\VEV{B_0}$ from
the second one.}   

Because the dynamics in this model is under control for small
$g$, $g^{\prime}$ and $\lambda$, the model provides a proof
that the dynamics
with vector condensate is a real thing. Moreover, this dynamics is
quite rich. In particular, as was shown in Ref. \cite{Gorbar:2005pi},
there are three types of topologically stable vortices 
in model (\ref{Lagrangian}), which are connected
either with photon field or hypercharge gauge field, or 
with both of them. As we will see below, the dynamics in the gluonic
phase strikingly resembles the dynamics in this toy model being
however much more complicated.

\section{Gauged NJL model}
\label{sec3}

We study dense two-flavor quark matter in $\beta$-equilibrium.
For our purpose, it is convenient to use a phenomenological
NJL model with gluons, the gauged NJL model. As was already
pointed out in the Introduction, 
this approach corresponds to the Higgs picture.
The confinement picture will be considered in Subsec.\ref{sec8}.

For simplicity, the current quark masses and the 
$(\bar{\psi}\psi)^2$ interaction in chiral channels will
be neglected.
Then the Lagrangian density is given by
\begin{equation}
  {\cal L} = \bar{\psi}(i\fsl{D}+\hat{\mu}_0\gamma^0)\psi
  +G_\Delta \bigg[\,(\bar{\psi}^C i\varepsilon\epsilon^\alpha\gamma_5 \psi)
           (\bar{\psi} i\varepsilon\epsilon^\alpha\gamma_5 \psi^C)\,\bigg]
  -\frac{1}{4}F_{\mu\nu}^{(a)} F^{(a)\,\mu\nu} ,
 \label{Lag}
\end{equation}
where
\begin{equation}
  D_\mu \equiv \partial_\mu - i g A_\mu^{(a)} T^{a}, \qquad
  F_{\mu\nu}^{(a)}
  \equiv \partial_\mu A_\nu^{(a)} - \partial_\nu A_\mu^{(a)} +
  g f^{abc} A_\mu^{(b)} A_\nu^{(c)}.
\end{equation}
Here $A_\mu^{(a)}$ are gluon fields,
$T^{a}$ are the generators of $SU(3)$ in the fundamental representation,
and $f^{abc}$ are the structure constants of $SU(3)$.
The spinor field $\psi=\psi_{i\alpha}$ has the flavor ($i=u,d$) and
color ($\alpha=r, g, b$) indices 
($\psi_{ur}$ is an up-red quark, etc.).
In Eq.~(\ref{Lag}), $(\varepsilon) \equiv \epsilon^{ij}$,
($\epsilon^{ud} = +1$), and
$(\epsilon^{\alpha}) \equiv \epsilon^{\alpha\beta\gamma}$,
($\epsilon^{rgb}=+1$) are the totally antisymmetric tensors
in the flavor and color spaces, respectively.
Note that we do not write explicitly the free electron term, and
the photon field is not introduced in the model.

In the $\beta$-equilibrium, the chemical potential matrix $\hat{\mu}_0$
for up and down quarks is:
\begin{equation}
  \hat{\mu}_0 = \mu {\bf 1} - \mu_e Q_{\rm em} + \mu_8 Q_8\, , 
 \label{mu}
\end{equation}
where 
${\bf 1} \equiv {\bf 1}_c \otimes {\bf 1}_f$,
$Q_{\rm em} \equiv {\bf 1}_c \otimes \diag(2/3,-1/3)_f$,
$Q_8 \equiv \diag(1/3,1/3,-2/3)_c\otimes {\bf 1}_f$,
and 
$\mu$, $\mu_e$ and $\mu_8$
are the quark, electron and color chemical potentials,
respectively
(the baryon chemical potential $\mu_B$ is $\mu_B \equiv 3\mu$).
Here the subscripts $c$ and $f$ mean that the corresponding matrices act in
the color and flavor spaces, respectively.
[Henceforth we will not show explicitly the unit matrices ${\bf 1}$,
${\bf 1}_c$, and ${\bf 1}_f$.] Note that the status of the
chemical potential $\mu_8$ is somewhat different from that of
$\mu$ and $\mu_e$ (in the absence of a photon field). 
The point is that the color neutrality
condition is nothing else as the Gauss's law for the gluon field 
$A_0^{(8)}$ and $\mu_8$ is expressed through $A_0^{(8)}$  
as
\begin{equation}
\mu_8 \equiv \frac{\sqrt{3}}{2}g A_0^{(8)}.
\label{mu8}
\end{equation}
As was shown in Ref. \cite{Gerhold:2003js}, $\mu_8$ is nonzero
in the 2SC/g2SC phase. Although the color chemical potential is
not an independent quantity in the gauged NJL model, we will
keep the notation $\mu_8$ in order to exhibit a special role of the
field $A_0^{(8)}$.

Eq. (\ref{mu}) implies  
that the total chemical potentials for different quarks
in the 2SC/g2SC phase are
\begin{subequations}
\label{beta-eq}
\begin{align}
  \mu_{ur} &= \mu_{ug} = \tilde{\mu} - \delta \mu,
  &\mu_{ub} &= \tilde{\mu} - \delta \mu - \mu_8, \\
   \mu_{dr}&= \mu_{dg} = \tilde{\mu} + \delta \mu,
  &\mu_{db}&= \tilde{\mu} + \delta \mu - \mu_8 ,
\end{align}
\end{subequations}
with
\begin{equation}
  \tilde{\mu} \equiv \mu - \frac{\mu_e}{6}+\frac{\mu_8}{3} , \quad
  \delta \mu \equiv \frac{\mu_e}{2}.
\label{tildemu}
\end{equation}
Let us now introduce the diquark field
$\Phi^\alpha \sim i\bar{\psi}^C\varepsilon \epsilon^\alpha \gamma_5 \psi$.
Then one can rewrite the Lagrangian density (\ref{Lag}) as
\begin{equation}
  {\cal L} = \bar{\psi}(i\fsl{D}+\hat{\mu}_0 \gamma^0)\psi
  - \frac{|\Phi^\alpha|^2}{4G_\Delta}
  - \frac{1}{2}\Phi^\alpha
     [i\bar{\psi}\varepsilon\epsilon^\alpha \gamma_5 \psi^C]
 - \frac{1}{2}
    [i\bar{\psi}^C\varepsilon\epsilon^\alpha\gamma_5 \psi]\Phi^{*\alpha}
 - \frac{1}{4}F_{\mu\nu}^{(a)} F^{(a)\,\mu\nu} .
 \label{Lag_aux}
\end{equation}

Without loss of generality, 
the diquark condensate in the 2SC/g2SC phase
can be chosen along the anti-blue direction:
$\VEV{\Phi^r}=0, \quad  \VEV{\Phi^g}=0, \quad \VEV{\Phi^b}\neq 0$ .
In correspondence with that,  
in the gauged NJL model, it will be convenient to use 
the following (partly unitary) gauge in the 2SC/g2SC phase
\cite{footnote}:
\begin{equation}
  \Phi^r =0, \quad  \Phi^g=0, \quad \Phi^b\equiv\Delta 
  \label{2SC}
\end{equation}
with the field $\Delta$ being real. Then the gap $\bar{\Delta}$
in the 2SC/g2SC phase is equal to 
the vacuum expectation 
value of $\Delta$, $\bar{\Delta} \equiv \VEV{\Delta}$ .

We now introduce the Nambu-Gor'kov spinor,
\begin{equation}
  \Psi \equiv \left(\begin{array}{@{}c@{}} \psi \\ \psi^C \end{array}\right) .
\end{equation}
The inverse propagator of $\Psi$ with the field $\Delta$ in
the 2SC/g2SC phase is given by
\begin{equation}
  S^{-1}(P) = \left(
  \begin{array}{cc}
  [G_0^+]^{-1} & \Delta^- \\ \Delta^+ &  [G_0^-]^{-1}
  \end{array}
  \right)
  \label{S-inv}
\end{equation}
with
\begin{equation}
  [G_0^+]^{-1}(P) \equiv
  (p_0+\tilde{\mu}-\delta\mu\tau_3-\mu_8{\bf 1}_b)\gamma^0
  -\vec \gamma \cdot \vec p,
\end{equation}
\begin{equation}
  [G_0^-]^{-1}(P) \equiv
  (p_0-\tilde{\mu}+\delta\mu\tau_3+\mu_8{\bf 1}_b)\gamma^0
  -\vec \gamma \cdot \vec p,
\end{equation}
and
\begin{equation}
  \Delta^- \equiv -i\varepsilon\epsilon^b\gamma_5\Delta, \qquad
  \Delta^+ \equiv \gamma^0 (\Delta^{-})^\dagger \gamma^0 =
  -i\varepsilon\epsilon^b\gamma_5\Delta,
\end{equation}
where $P^\mu \equiv (p_0,\vec p)$ is an energy-momentum four vector,
$\tau_3 \equiv \diag(1,-1)_f$ and
${\bf 1}_b \equiv \diag (0,0,1)_c$.

The propagator of $\Psi$ is given by
\begin{equation}
  S(P) = \left(
  \begin{array}{cc}
  G^+ & \Xi^- \\ \Xi^+ &  G^-
  \end{array}
  \right) 
  \label{S}
\end{equation}
with
\begin{equation}
  G^\pm = \{[G_0^\pm]^{-1} - \Delta^\mp G_0^\mp \Delta^\pm\}^{-1}, \qquad
  \Xi^\pm = - G_0^\mp \Delta^\pm G^\pm .
  \label{G-Xi}
\end{equation}
The structure of $G^\pm$ and $\Xi^\pm$ was determined in  
the second paper in Ref.~\cite{Huang:2004bg}:
\begin{equation}
  G^\pm(P) \equiv \diag (G_\Delta^\pm,G_\Delta^\pm,G_b^\pm)_c,
\end{equation}
where
\begin{eqnarray}
  G_\Delta^\pm(P) &=& \phantom{+}
  \frac{(p_0\mp\delta\mu\tau_3)-E^\pm}
       {(p_0\mp\delta\mu\tau_3)^2-(E^\pm_\Delta)^2}\gamma^0 \Lambda_p^+
 +\frac{(p_0\mp\delta\mu\tau_3)+E^\mp}
       {(p_0\mp\delta\mu\tau_3)^2-(E^\mp_\Delta)^2}\gamma^0 \Lambda_p^-,
  \label{G_del} \\[3mm]
  G_b^\pm(P) &=& \phantom{+}
  \frac{1}{(p_0\mp\delta\mu\tau_3\mp\mu_8)+E^\pm}\gamma^0 \Lambda_p^+
 +\frac{1}{(p_0\mp\delta\mu\tau_3\mp\mu_8)-E^\mp}\gamma^0 \Lambda_p^-,
  \label{G_b}
\end{eqnarray}
and
\begin{equation}
  \Xi^\pm(P) \equiv \epsilon^b
  \left(\begin{array}{cc}
    0 & \Xi_{12}^\pm \\ -\Xi_{21}^\pm & 0
  \end{array}\right)_f
\end{equation}
with
\begin{eqnarray}
  \Xi_{12}^\pm (P) &=&
  -i\Delta\left[\,\frac{1}{(p_0 \pm \delta\mu)^2-(E^\pm_\Delta)^2}
                   \gamma_5\Lambda_p^+ 
                 +\frac{1}{(p_0 \pm \delta\mu)^2-(E^\mp_\Delta)^2}
                   \gamma_5\Lambda_p^-\,\right], \\
  \Xi_{21}^\pm (P) &=&
  -i\Delta\left[\,\frac{1}{(p_0 \mp \delta\mu)^2-(E^\pm_\Delta)^2}
                   \gamma_5\Lambda_p^+ 
                 +\frac{1}{(p_0 \mp \delta\mu)^2-(E^\mp_\Delta)^2}
                   \gamma_5\Lambda_p^-\,\right].
  \label{Xi}
\end{eqnarray}
Here
\begin{equation}
   E^\pm \equiv |\vec p| \pm \tilde{\mu}, \qquad
   E^\pm_\Delta \equiv \sqrt{(E^\pm)^2 + \Delta^2}, \qquad
  \Lambda_p^\pm \equiv \frac{1}{2}
  \left(1\pm\gamma^0\frac{\vec \gamma \cdot \vec p}{|\vec p|}\right),
\end{equation}
and while $G_{\Delta}^\pm$ and $G_b^\pm$ are $8 \times 8$ matrices in
the flavor-spinor space, the $4 \times 4$ matrices
$\Xi_{12}^\pm$ and $\Xi_{21}^\pm$ act only in the spinor space. 

The generalization of expression (\ref{S-inv}) for the inverse
propagator of $\Psi$ both with the scalar
diquark field $\Delta$ and the vector fields $A_\mu^{(a)}$
in a gluonic phase (with nonzero gluonic
condensates  $\VEV{A_\mu^{(a)}}$ ) is straightforward:
\begin{equation}
  S_g^{-1}(P) = \left(
  \begin{array}{cc}
  [G_{0,g}^+]^{-1} & \Delta^- \\ \Delta^+ &  [G_{0,g}^-]^{-1}
  \end{array}
  \right) ,
  \label{Sg-inv}
\end{equation}
with
\begin{equation}
  [G_{0,g}^+]^{-1}(P) \equiv
  (p_0+\hat{\mu}_0)\gamma^0
  -\vec \gamma \cdot \vec p + g \fsl{A}^{(a)}T^a,
\end{equation}
\begin{equation}
  [G_{0,g}^-]^{-1}(P) \equiv
  (p_0-\hat{\mu}_0)\gamma^0
  -\vec \gamma \cdot \vec p - g \fsl{A}^{(a)}(T^a)^T .
\end{equation}

Integrating out fermion fields, we obtain
the potential including both gluon and diquark fields: 
\begin{equation}
  V =
   \frac{\Delta^2}{4G_\Delta}
  +\frac{g^2}{4}f^{a_1a_2a_3}f^{a_1a_4a_5}
    A_\mu^{(a_2)} A_\nu^{(a_3)} A^{(a_4)\,\mu} 
A^{(a_5)\,\nu}
  -\frac{1}{2}\int\frac{d^4 P}{i(2\pi)^4}\Tr\ln S_g^{-1} .
  \label{V}
\end{equation}
We will utilize the hard dense loop approximation, in which only
the dominant one-loop quark contribution is taken into account,
while the contribution of gluon loops is neglected. On the other hand,
we keep the tree contribution of gluons in the 
effective potential (\ref{V}). This is because we want to compare 
this contribution with that of hard dense loops in order to check the
consistency of the hard dense loop approximation. The ground state
of the system corresponds to the minimum of potential (\ref{V})
and the question whether there exist gluon condensates  
$\VEV{A^{(a)}_\mu}$ is a dynamical issue.

The following remarks are in order. 

a) To study the ground state in the Higgs phase, it
is convenient to use the unitary gauge. The important point is
that in this gauge, all
auxiliary (gauge dependent) degrees of freedom are removed.
{\it Therefore in this gauge the vacuum expectations values (VEVs)
$\VEV{A^{(a)}_\mu}$ of vector fields are well-defined
physical quantities.} The unitary gauge in the gluonic phase,
which uses and extends the constraints presented in Eq. (\ref{2SC}),
will be described in Sec. \ref{gluon}.

b) As we will see below, in the gluonic phase,
the time-component VEVs of the gluon fields other than the 8th one
are also nonzero. Because of that,
it will be convenient to rewrite effective potential
(\ref{V}) in a somewhat different form.
Let us introduce the following 
matrix ${\cal M}_g$ in the Nambu-Gor'kov space,
\begin{equation}
  {\cal M}_g \equiv S^{-1}_g - S^{-1} =
  \left(
  \begin{array}{cc}
  \mu_{\breve{a}} T^{\breve{a}} \gamma^0
  - g\vec A^a \cdot \vec \gamma  \;T^a & 0 \\
  0 & -\mu_{\breve{a}} (T^{\breve{a}})^T \gamma^0
  + g\vec A^a \cdot \vec \gamma \;(T^a)^T
  \end{array}
  \right) ,
\end{equation}
where 
\begin{equation}
  \mu_{\breve{a}} \equiv g A_0^{\breve{a}},
\quad (\breve{a}=1,2,\cdots,7)
\end{equation}
($\mu_8\equiv \frac{\sqrt{3}}{2}gA_0^{(8)}$ is included in $S^{-1}$).
Expanding now the logarithmic term in Eq.~(\ref{V}),
we find
\begin{equation}
  V = V_\Delta (\Delta,\mu_e,\mu_8)
  +\frac{g^2}{4}f^{a_1a_2a_3}f^{a_1a_4a_5}
    A_\mu^{(a_2)} A_\nu^{(a_3)} A^{(a_4)\,\mu} A^{(a_5)\,\nu}
  +\sum_{n=1}^\infty \frac{(-1)^n}{2n}\int\frac{d^4 P}{i(2\pi)^4}
   \Tr (S{\cal M}_g)^n,
  \label{V_exp}
\end{equation}
where we defined the 2SC/g2SC part of the effective potential as
\begin{equation}
  V_\Delta (\Delta,\mu_e,\mu_8) = \frac{\Delta^2}{4G_\Delta}
 -\frac{1}{2}\int\frac{d^4 P}{i(2\pi)^4}\Tr\ln S^{-1} .
\label{Vd}
\end{equation}
The form (\ref{V_exp})
of the effective potential will be used in our
analysis below.

\section{$SU(2)_c$ decomposition}
\label{sec4}

 Because the diquark condensate 
$\bar{\Delta} = \VEV{\Phi^b}\equiv \VEV{\Delta}$
breaks the initial color
$SU(3)_c$ symmetry down to the $SU(2)_c$ one, it is useful to decompose
the initial fields with respect to the  $SU(2)_c$ subgroup.
In particular,
the decomposition will help to calculate systematically
the effective potential (\ref{V_exp}).

The (anti-) fundamental and adjoint representations of $SU(3)_c$ are
decomposed with respect to the $SU(2)_c$ as:
\begin{equation}
  {\bf 3}={\bf 2}\oplus{\bf 1}, \quad \mbox{i.e.,} \quad
  \left(
    \begin{array}{@{}c@{}}
      \psi_{ir} \\ \psi_{ig} \\ \psi_{ib}
    \end{array}
  \right) =
  \left(
    \begin{array}{@{}c@{}}
      \psi_{ir} \\ \psi_{ig}
    \end{array}
  \right) \oplus \psi_{ib} ,  \qquad (i=u,d),
\end{equation}
\begin{equation}
  \bar{{\bf 3}}=\bar{{\bf 2}} \oplus{\bf 1}, \quad \mbox{i.e.,} \quad
  \left(
    \begin{array}{@{}c@{}}
      \Phi^r \\ \Phi^g \\ \Phi^b
    \end{array}
  \right) =
  \left(
    \begin{array}{@{}c@{}}
      \Phi^r \\ \Phi^g
    \end{array}
  \right) \oplus \Phi^b ,
\end{equation}
and
\begin{equation}
  {\bf 8}={\bf 3}\oplus{\bf 2}\oplus\bar{{\bf 2}}\oplus{\bf 1},
  \quad \mbox{i.e.,} \quad
 \{A_\mu^{a}\} = (A_\mu^{(1)},A_\mu^{(2)},A_\mu^{(3)})
 \oplus \phi_\mu \oplus \phi_\mu^* \oplus A_\mu^{(8)} ,
 \quad (a=1,2,\cdots,8).
\end{equation}
Here we defined the complex doublets of the matter 
(with respect to the $SU(2)_c$) vector fields,
\begin{equation}
  \phi_\mu \equiv
  \left(
  \begin{array}{@{}c@{}} \phi_\mu^r \\[2mm] \phi_\mu^g \end{array}
  \right)
  =
 \frac{1}{\sqrt{2}}
  \left(
  \begin{array}{c}
  A_\mu^{(4)}-iA_\mu^{(5)} \\[2mm] A_\mu^{(6)}-iA_\mu^{(7)}
  \end{array}
  \right) , \qquad
  \phi_\mu^* \equiv
  \left(
  \begin{array}{@{}c@{}} \phi_\mu^{*r} \\[2mm] \phi_\mu^{*g} \end{array}
  \right)
  =
 \frac{1}{\sqrt{2}}
  \left(
  \begin{array}{c}
  A_\mu^{(4)}+iA_\mu^{(5)} \\[2mm] A_\mu^{(6)}+iA_\mu^{(7)}
  \end{array}
  \right) .
\label{phi}
\end{equation}

Then we define the field strength for the $SU(2)_c$ 
gauge bosons,
\begin{equation}
  f_{\mu\nu}^{(\ell)} \equiv \partial_\mu A_\nu^{(\ell)} -
  \partial_\nu A_\mu^{(\ell)}
  + g\epsilon^{\ell m n} A_\mu^{(m)} A_\nu^{(n)} \qquad
  (\ell, m, n=1,2,3)\, ,
\label{su2strength}
\end{equation}
and the covariant derivative
\begin{equation}
  {\cal D}_\mu \equiv \partial_\mu - i g A_\mu^{(\ell)}
  \frac{\sigma^\ell}{2} \quad (\ell = 1,2,3) \, .
\end{equation}
It will be also useful to define the combinations
\begin{equation}
  {\cal A}_\mu^+ \equiv \frac{1}{\sqrt{2}}(A_\mu^{(1)}+iA_\mu^{(2)}), \quad
  {\cal A}_\mu^- \equiv \frac{1}{\sqrt{2}}(A_\mu^{(1)}-iA_\mu^{(2)}).
\end{equation}

Because of the presence of the electric chemical potential
$\mu_e$ in the model, 
the chiral $SU(2)_{L,R}$ symmetry is explicitly broken down to
its $U(1)$-part $U(1)_{\tau^3_{L,R}}$.
Therefore the initial symmetry in the gauged NJL model is
\begin{equation}
 SU(3)_c \times U(1)_{\rm em} \times
 [U(1)_{\tau^3_L} \times U(1)_{\tau^3_R}]_{\chi} \times SO(3)_{\rm rot}.
 \label{sym-init}
\end{equation}
Since a photon filed was not included,
the electromagnetic symmetry $U(1)_{\rm em}$ is global.
Note that the initial baryon charge is 
\begin{equation}
  {\cal B} = \frac{1}{3} {\bf 1}_f \otimes {\bf 1}_c = 2(Q_{\rm em}-I_3),
\end{equation}
where $I_3$ is a diagonal subgroup of the
$U(1)_{\tau^3_L} \times U(1)_{\tau^3_R}$, i.e., $I_3=\diag(1/2,-1/2)_f$.

The diquark gap $\bar{\Delta}$ breaks the initial symmetry 
(\ref{sym-init}) down to
\begin{equation}
 SU(2)_c \times
 \tilde{U}(1)_{\rm em} \times
 [U(1)_{\tau^3_L} \times U(1)_{\tau^3_R}]_{\chi} \times SO(3)_{\rm rot}.
 \label{sym-2SC}
\end{equation}
The unbroken $\tilde{U}(1)_{\rm em}$ is connected with  
the new electric charge 
\begin{equation}
  \tilde{Q}_{\rm em} = Q_{\rm em}-\frac{1}{\sqrt{3}}T^8,
\end{equation}
where $T^8 \equiv \frac{1}{2\sqrt{3}} \diag(1,1,-2)_c$.
The new unbroken baryon charge is 
\begin{equation}
  \tilde{\cal B} = 2(\tilde{Q}_{\rm em} - I_3) .
\label{newB}
\end{equation}
The charges for fermions, diquark fields and gluons are summarized in
Tables \ref{tab1} and \ref{tab2}.

The transformations of the fields under the gauge $SU(2)_c$ 
have the following form
\begin{eqnarray}
  \left(
    \begin{array}{@{}c@{}}
      \psi_{ir} \\ \psi_{ig}
    \end{array}
  \right) &\to&
  \left(
    \begin{array}{@{}c@{}}
      \psi'_{ir} \\ \psi'_{ig}
    \end{array}
  \right) = U
  \left(
    \begin{array}{@{}c@{}}
      \psi_{ir} \\ \psi_{ig}
    \end{array}
  \right), \qquad (i=u,d), \\[3mm]
  \left(
    \begin{array}{@{}c@{}}
      \Phi^r \\ \Phi^g
    \end{array}
  \right) &\to&
  \left(
    \begin{array}{@{}c@{}}
      \Phi'{}^r \\ \Phi'{}^g
    \end{array}
  \right) = U^*
  \left(
    \begin{array}{@{}c@{}}
      \Phi^r \\ \Phi^g
    \end{array}
  \right), \\[3mm]
  \phi_\mu &\to& \phi_\mu'= U \phi_\mu, \\[3mm]
  {\cal A}_\mu &\to& {\cal A}'_\mu = U {\cal A}_\mu U^{-1}
  +\frac{i}{g}U \partial_\mu U^{-1}, \\[3mm]
  f_{\mu\nu} &\to& f'_{\mu\nu} = U f_{\mu\nu} U^{-1}, \\[3mm]
  {\cal D}_\mu &\to& {\cal D}'_\mu = U {\cal D}_\mu U^{-1} ,
\end{eqnarray}
where
\begin{equation}
  {\cal A}_\mu \equiv A_\mu^{(\ell)} \frac{\sigma^\ell}{2}, \quad
  f_{\mu\nu} \equiv f_{\mu\nu}^{(\ell)} \frac{\sigma^\ell}{2},
  \quad (\ell = 1,2,3)
\end{equation}
and
\begin{equation}
  U=\exp\left(i\theta^\ell(x)\frac{\sigma^\ell}{2}\right) .
\end{equation}
Under the new baryon symmetry connected with generator
$\tilde{\cal B}$ (\ref{newB}),
blue quarks, the diquark fields $\Phi^{r,g}$ and the $\phi_\mu$ field doublet
transform as
\begin{equation}
 \psi_{ib} \to \psi'_{ib}=e^{i\theta}\psi_{ib}, \; (i=u,d), \quad
 \Phi^{r,g} \to \Phi'{}^{r,g}=e^{i\theta}\Phi^{r,g}, \quad
 \phi_\mu \to \phi_\mu'= e^{-i\theta} \phi_\mu ,
\end{equation}
while other fields have zero baryon charge $\tilde{\cal B}$
(see  Tables \ref{tab1} and \ref{tab2}.)

\renewcommand{\arraystretch}{2.0}
\begin{table}
  \begin{tabular}{|c|cc|cc|cc|c|}\hline
          & $Q_{\rm em}$ & ${\cal B}$ & $\tilde{Q}_{\rm em}$ & $\tilde{\cal B}$
          & $\tilde{\tilde{Q}}_{\rm em}$ & $\tilde{\tilde{\cal B}}$ & $I_3$ \\
    \hline \hline
    $\psi_{ur}$ & $\frac{2}{3}$ & $\frac{1}{3}$ & $\frac{1}{2}$ & 0
          & 0 & $-1$ & $\frac{1}{2}$ \\
    $\psi_{ug}$ & $\frac{2}{3}$ & $\frac{1}{3}$ & $\frac{1}{2}$ & 0
          & 1 & $1$ & $\frac{1}{2}$ \\
    $\psi_{ub}$ & $\frac{2}{3}$ & $\frac{1}{3}$ & 1 & 1
          & 1 & 1 & $\frac{1}{2}$ \\ \hline\hline
    $\psi_{dr}$ & $-\frac{1}{3}$ & $\frac{1}{3}$ & $-\frac{1}{2}$ & 0
          & $-1$ & $-1$ & $-\frac{1}{2}$ \\
    $\psi_{dg}$ & $-\frac{1}{3}$ & $\frac{1}{3}$ & $-\frac{1}{2}$ & 0
          & $0$ & $1$ & $-\frac{1}{2}$ \\
    $\psi_{db}$ & $-\frac{1}{3}$ & $\frac{1}{3}$ & 0 & 1
          & $0$ & $1$ & $-\frac{1}{2}$ \\ \hline\hline
    $\Phi^r$ & $\frac{1}{3}$ & $\frac{2}{3}$ & $\frac{1}{2}$ & $1$
          & $1$ & $2$ & $0$ \\
    $\Phi^g$ & $\frac{1}{3}$ & $\frac{2}{3}$ & $\frac{1}{2}$ & $1$
          & $0$ & $0$ & $0$ \\
    $\Phi^b$ & $\frac{1}{3}$ & $\frac{2}{3}$ & 0 & 0
          & $0$ & $0$ & $0$ \\ \hline
  \end{tabular}
  \caption{The quantum numbers of the up and down quarks and
           the diquark fields. In the unitary gauge the diquark fields
           $\Phi^{r,g}$ and $\mbox{Im}\Phi^b$ are absorbed into
           the longitudinal modes of the corresponding gluons.
           \label{tab1}}
\end{table}

\begin{table}
  \begin{tabular}{|c|cc|cc|cc|c|}\hline
          & $Q_{\rm em}$ & ${\cal B}$ & $\tilde{Q}_{\rm em}$ & $\tilde{\cal B}$
          & $\tilde{\tilde{Q}}_{\rm em}$ & $\tilde{\tilde{\cal B}}$ & $I_3$ \\
    \hline \hline
    ${\cal A}_\mu^+ \equiv \frac{1}{\sqrt{2}}(A_\mu^{(1)}+iA_\mu^{(2)})$
    & 0 & 0 & 0 & 0 & $1$ & $2$ & 0 \\
    $A_\mu^{(3)}$ & 0 & 0 & 0 & 0 & $0$ & $0$ & 0 \\
    ${\cal A}_\mu^- \equiv \frac{1}{\sqrt{2}}(A_\mu^{(1)}-iA_\mu^{(2)})$
    & 0 & 0 & 0 & 0 & $-1$ & $-2$ & 0 \\ \hline\hline
    $\phi_\mu^{*r} \equiv \frac{1}{\sqrt{2}}(A_\mu^{(4)}+iA_\mu^{(5)})$
    & 0 & 0 & $\frac{1}{2}$ & $1$ & $1$ & $2$ & 0 \\
    $\phi_\mu^{*g} \equiv \frac{1}{\sqrt{2}}(A_\mu^{(6)}+iA_\mu^{(7)})$
    & 0 & 0 & $\frac{1}{2}$ & $1$ & $0$ & $0$ & 0 \\ \hline\hline
    $\phi_\mu^r \equiv \frac{1}{\sqrt{2}}(A_\mu^{(4)}-iA_\mu^{(5)})$
    & 0 & 0 & $-\frac{1}{2}$ & $-1$ & $-1$ & $-2$ & 0 \\
    $\phi_\mu^g \equiv \frac{1}{\sqrt{2}}(A_\mu^{(6)}-iA_\mu^{(7)})$
    & 0 & 0 & $-\frac{1}{2}$ & $-1$ & $0$ & $0$ & 0 \\ \hline\hline
    $A_\mu^{(8)}$ & 0 & 0 & 0 & 0 & $0$ & $0$ & 0 \\ \hline
  \end{tabular}
  \caption{The quantum numbers of gluons.\label{tab2}}
\end{table}
\renewcommand{\arraystretch}{1.0}

\section{Gluonic phase}
\label{gluon}

Both the chromomagnetic \cite{Huang:2004bg} and plasmon 
\cite{Gorbar:2006up} instabilities for the 4-7th gluons
in the 2SC phase at $\bar{\Delta} < \sqrt{2}\delta\mu$
suggest a condensation of these gluons,
i.e., the creation of a condensate
of vector field $\phi_\mu$ (\ref{phi}). 
Because the chromomagnetic instability develops in the magnetic
channel, 
it is naturally to expect that
a spatial component of $\phi_\mu$ has a VEV. In studying this
condensate, one can use 
the freedom connected with symmetry (\ref{sym-2SC}) in the 2SC/g2SC phase.
By using the rotational symmetry $SO(3)_{\rm rot}$, one can take
$\VEV{\phi_{z}} \ne 0$ while 
$\VEV{\phi_{x}}=\VEV{\phi_{y}}=0$. 
And because of the $SU(2)_c$ symmetry,
without loss of generality, we can choose $\VEV{A^{(6)}_{z}} \ne 0$.
This VEV breaks the $SU(2)_c$ down to nothing  
and the $SO(3)_{\rm rot}$ down to the $SO(2)_{\rm rot}$.

The following remarks are in order. a) The complex doublet $\phi_{z}$ plays
here the role of a Higgs field responsible for spontaneous
breakdown of the $SU(2)_c$. The situation is similar to that
taking place in the electroweak theory. The essential difference
however is that now the Higgs field is a spatial component of the
vector field leading also to spontaneous breakdown of the
rotational symmetry. b) In this paper, we will use the gauge in which
\begin{equation}
\phi_{z}^T = \frac{1}{\sqrt{2}}(0, \VEV{A^{(6)}_{z}} + a^{(6)}_{z}),
\label{unitary}
\end{equation}
where the real field $a^{(6)}_{z}$ describes quantum fluctuations. 
This constraint together with that in Eq. (\ref{2SC}) 
constitute the unitary gauge: all auxiliary (gauge dependent) degrees of
freedom are now removed.

A gluonic phase with such
a condensate was described in letter \cite{Gorbar:2005rx}.
Since the most of the initial symmetries are broken in
this phase, its dynamics is rich and complicated. On the
other hand, because of space shortage, the description of
this phase in letter \cite{Gorbar:2005rx} was rather
brief. In this paper,
we will present both a detailed 
description of its dynamics and present a general analysis
of a possibility of the existence of other phases with
vector gluon condensates in dense QCD. 

\subsection{Symmetry breaking structure
and Ginzburg-Landau effective potential}
\label{sec5}

Let us describe symmetry breaking structure
in the gluonic phase. With a broken $SU(2)_c$, the $SU(2)_c$ gluons 
could have 
VEVs. A similar situation takes place in the gauged $\sigma$-model
with a chemical potential for hypercharge described in Sec. \ref{sec2}
above: the gauge symmetry
$SU(2)_L$ is broken there. 
Motivating by that model, we assume
\begin{equation}
  \VEV{A^{(1)}_{z}}, \;\VEV{A^{(3)}_{0}} \ne 0.
\end{equation}
As will be shown below, a
solution with these vector condensates
exists in the model indeed.

The symmetry in the 2SC/g2SC phase is that presented in 
Eq. (\ref{sym-2SC}). The VEV $\VEV{A^{(6)}_{z}}$ breaks the $SU(2)_c$ 
but a linear combination of the generator $T^3$ from the
$SU(2)_c$ and $\tilde{Q}_{\rm em}$,
\begin{equation}
  \tilde{\tilde{Q}}_{\rm em} = \tilde{Q}_{\rm em} - T^3 = Q_{\rm em}
-\frac{1}{\sqrt{3}}T^8 - T^3,
\end{equation}
determines the unbroken $\tilde{\tilde{U}}(1)_{\rm em}$. The new baryon
charge is
${\tilde{\tilde{\cal B}}} = 2(\tilde{\tilde{Q}}_{\rm em} - I_3)$
[the charges $\tilde{\tilde{Q}}_{\rm em}$ 
and ${\tilde{\tilde{\cal B}}}$ 
for quarks, diquarks, and gluons are shown in  
Tables \ref{tab1} and \ref{tab2}].
However, because $T^1$ does not commute with $T^3$,
the VEV $\VEV{A^{(1)}_{z}}$ breaks $\tilde{\tilde{U}}_{\rm em}(1)$. The
$U(1)$ symmetry connected with the
baryon charge ${\tilde{\tilde{\cal B}}}$ is also broken.

After all, we have:
\begin{eqnarray}
\lefteqn{\hspace*{-2cm}
 [SU(3)_c]_{\rm local} \times [U(1)_{\rm em} \times 
 U(1)_{\tau^3_L} \times U(1)_{\tau^3_R}]_{\rm global} \times SO(3)_{\rm rot} 
} \nonumber \\
&& \stackrel{\bar{\Delta}}{\longrightarrow}
 [SU(2)_c]_{\rm local} \times [\tilde{U}(1)_{\rm em} \times 
 U(1)_{\tau^3_L} \times U(1)_{\tau^3_R}]_{\rm global} \times SO(3)_{\rm rot}\\
&& \stackrel{\VEV{A^{(6)}_{z}}}{\longrightarrow}
 [\tilde{\tilde{U}}(1)_{\rm em} \times 
 U(1)_{\tau^3_L} \times U(1)_{\tau^3_R}]_{\rm global} \times SO(2)_{\rm rot}\\
&& \stackrel{\VEV{A^{(1)}_{z}}}{\longrightarrow}
 [U(1)_{\tau^3_L} \times U(1)_{\tau^3_R}]_{\rm global} \times SO(2)_{\rm rot} .
\end{eqnarray}
Thus, this system describes an anisotropic medium in which both the
color and electric superconductivities coexist.

Let us apply the Ginzburg-Landau
(GL) approach to this system near the critical point
$\delta\mu \simeq \bar{\Delta}/\sqrt{2}$.
The two point function of gluons can be calculated from 
Lagrangian density (\ref{Lag_aux}).
While in Ref.~\cite{Huang:2004bg}
the Debye and Meissner screening masses of the gluons in 
the 2SC phase were calculated, the pole masses
of the corresponding light plasmons (with masses $|M| \ll |\mu|$) 
were analyzed in 
Ref. \cite{Gorbar:2006up}.
For the gluons of the unbroken $SU(2)_c$, i.e., 
$A_{\mu}^{(1)}$, $A_{\mu}^{(2)}$, and $A_{\mu}^{(3)}$, both Debye and Meissner 
masses  
vanish in the region $\delta\mu < \bar{\Delta}$ and there are no light
plasmons in these channels.
For the gluons $A_{\mu}^{(4)-(7)}$, the Meissner mass is approximately
\begin{equation}
 m_{M,4}^2 = \frac{g^2\tilde{\mu}^2}{6\pi^2}
 \left(\,1-\frac{2\delta\mu^2}{\bar{\Delta}^2}\right), 
\quad \delta\mu < \bar{\Delta}.
\label{m_M4}
\end{equation}
Thus, near the critical point $\delta\mu=\bar{\Delta}/\sqrt{2}$,
the Meissner mass for $A_{\mu}^{(4)-(7)}$ is very small. As $\delta\mu$
exceeds the value $\bar{\Delta}/\sqrt{2}$,  $m_{M,4}^2$ becomes negative,
thus signalizing a chromomagnetic instability in the 2SC solution. 
The pole masses of the light plasmons for the magnetic and electric modes
have similar behavior in these channels.
On the other hand, around the critical point 
$\delta\mu=\bar{\Delta}/\sqrt{2}$,
the $SU(2)_c$ singlet gluon $A_{\mu}^{(8)}$ is heavy.
Actually, it is heavy in the whole region 
$\delta\mu < \bar{\Delta}$.
This fact allows us to pick up the gluons $A_{\mu}^{(1)-(7)}$ as relevant light
degrees of freedom in the low energy effective theory around 
the critical point $\delta\mu = \bar{\Delta}/\sqrt{2}$. 

Because the $SU(2)_c$ is a gauge symmetry,
the building blocks of the GL effective action are
\begin{equation}
  \phi_0, \quad \phi_j, \quad {\cal D}_0, \quad {\cal D}_j,  \quad
  f_{0j}, \quad f_{jk}, 
\label{GLblocks}
\end{equation}
where the indices $j$ and $k$ represent spatial components.
The $SU(2)_c$ and
$SO(3)_{\rm rot}$ 
symmetries dictate the form of the
general GL effective potential, which is made from these building blocks
and
includes operators up to the mass dimension four, i.e., relevant
and marginal ones.
It will be convenient to introduce the following notations:
\begin{equation}
B \equiv g A^{(6)}_{z}, \quad
C \equiv g A^{(1)}_{z}, \quad
D \equiv g A^{(3)}_{0}. 
\label{BCD}
\end{equation}
Then, as
shown in detail in Subsec. \ref{sec10} below, the GL potential has the form
\begin{equation}
  V_{\rm eff} = V_\Delta + \frac{1}{2}M_B^2 B^2
  + T_{DB} D B^2 + \frac{1}{2}\lambda_{BC} B^2 C^2
  + \frac{1}{2}\lambda_{BD} B^2 D^2
  + \frac{1}{2}\lambda_{CD} C^2 D^2 
  + \frac{1}{4}\lambda_B B^4,
  \label{LG-pot}
\end{equation}
where $V_\Delta$ is the 2SC part of the effective potential
(see Eq. (\ref{Vd})).
Here, while the coefficients $\lambda_B$,
$\lambda_{BC}$, $\lambda_{BD}$, and $\lambda_{CD}$ are dimensionless,
the dimension (in mass units) of the coefficient 
$T_{DB}$ in the triple vertex  
is one. Expanding the potential $V$ (\ref{V_exp}) with respect to $B$, $C$, 
and $D$, we can determine these coefficients.

Before realizing explicit calculations,
we clarify the behavior of the effective potential
(\ref{LG-pot}) near the critical point.
The stationary point of the effective potential (\ref{LG-pot})
is given by the equations
\begin{eqnarray}
 \frac{\partial V_{\rm eff}}{\partial B} &=&
 B \left[\,M_B^2+\lambda_B B^2+2T_{DB} D + \lambda_{BC}C^2 +
           \lambda_{BD} D^2\,\right] = 0,
 \label{gap-eq-B} \\
 \frac{\partial V_{\rm eff}}{\partial C} &=&
 C \left[\,\lambda_{BC}B^2 + \lambda_{CD} D^2\,\right] = 0,  
 \label{gap-eq-C} \\
 \frac{\partial V_{\rm eff}}{\partial D} &=&
 T_{DB} B^2 + \lambda_{BD} D B^2 + \lambda_{CD} C^2 D =0 ,
 \label{gap-eq-D}
\end{eqnarray}
and 
\begin{equation}
  \frac{\partial V_{\rm eff}}{\partial \mu_e} = 0, \qquad
  \frac{\partial V_{\rm eff}}{\partial \mu_8} = 0 , \qquad
  \frac{\partial V_{\rm eff}}{\partial \Delta} = 0. 
\label{parameters}
\end{equation}

It will be convenient to present $\mu_e$, $\mu_8$, and $\Delta$ as
\begin{eqnarray}
\label{bar1}
  \mu_e &=& \bar{\mu}_e + \xi_e, \\
  \mu_8 &=& \bar{\mu}_8 + \xi_8 , \\ 
  \Delta &=& \bar{\Delta} + \xi_\Delta,
\label{bar3}
\end{eqnarray}
where the bar-quantities are from the 2SC solution, with $B=C=D=0$.
Let us assume that the origin (bifurcation point)
of the solution with nonzero
$B$, $C$, and $D$ corresponds to a second order phase transition 
(as will become clear below, this assumption is self-consistent).
Under this assumption, the analysis of 
Eqs. (\ref{gap-eq-B})--(\ref{gap-eq-D}) and (\ref{parameters})
was done in Appendix \ref{A}.  
Taking an infinitesimally small $B$ near the critical point,
it is shown there that
\begin{equation}
  \xi_e, \xi_8, \xi_\Delta \sim {\cal O}(B^2)
  \label{xi}
\end{equation}
and that when the 2SC solution becomes unstable ($M_B^2 < 0$),
a new solution occurs, {\sl if} the parameters $\lambda_{BC}$ and
$\lambda_{CD}$ satisfy
\begin{equation}
  \lambda_{BC} > 0, \quad \lambda_{CD} < 0
\label{constraint}
\end{equation}
(in the next section, it will be shown that this constraint is
satisfied indeed). The new solution is:
\begin{equation}
  B_{\rm sol} \equiv g\VEV{A^{(6)}_z}\simeq \frac{-M_B^2}{3|T_{DB}|}
           \sqrt{\frac{-\lambda_{CD}}{\lambda_{BC}}}, \quad
  C_{\rm sol} \equiv g\VEV{A^{(1)}_z}
\simeq  \sqrt{\frac{-M_B^2}{3\lambda_{BC}}}, \quad
  D_{\rm sol} \equiv g\VEV{A^{(3)}_0}
\simeq \frac{-M_B^2}{3T_{DB}},
  \label{appr}
\end{equation}
where we neglected higher order terms in $M_B^2$. It is
important that, as shown in Appendix \ref{A},
the coefficients $M_B^2$, $T_{DB}$,
$\lambda_{BC}$, and $\lambda_{CD}$ in this nearcritical solution
are expressed through
the 2SC values $\bar{\Delta}$, $\bar{\mu}_e$ and $\bar{\mu}_8$.
Note that in Eq.~(\ref{appr}) the convention $B > 0$ and $C > 0$ 
is chosen.

Near the critical point $M_B^2 = 0$, 
the solution behaves as
\begin{equation}
  B_{\rm sol} \propto -M_B^2, \quad C_{\rm sol} \propto \sqrt{-M_B^2}, \quad
  D_{\rm sol} \propto -M_B^2 .
  \label{scaling}
\end{equation}
These scaling relations are quite remarkable.
While the scaling relation for $C$ is of engineering type, 
those for $B$ and $D$ are not (the origin of this is of
course in the presence of the 
dimensional coefficient $T_{DB}$ in Eq. (\ref{appr})).
Such a scaling behavior implies that the $B^4$ and $B^2 D^2$ terms  
in the effective potential are irrelevant
near the critical point $M_B^2 =0$. Omitting them,
we arrive at the reduced effective potential:
\begin{equation}
  \tilde{V}_{\rm eff} =
  V_\Delta + \frac{1}{2}M_B^2 B^2
  + T_{DB} D B^2 + \frac{1}{2}\lambda_{BC} B^2 C^2
  + \frac{1}{2}\lambda_{CD} C^2 D^2 .
  \label{V_min}
\end{equation}

Let us now turn to the 2SC part $V_\Delta$ in Eq. (\ref{V_min}). As shown
in Appendix \ref{A}, the difference of $V_\Delta$ in  
the new solution and that in the 2SC one is
\begin{equation}
   V_\Delta(\Delta^{\rm sol}, \mu_e^{\rm sol},\mu_8^{\rm sol})
 - V_\Delta(\bar{\Delta}, \bar{\mu}_e,\bar{\mu}_8) \sim {\cal O}(B^4) .
 \label{diff}
\end{equation}
On the other hand, as follows from Eqs. (\ref{scaling})
and (\ref{V_min}), the difference 
$\tilde{V}_{\rm eff}-V_\Delta \sim {\cal O}(B^3)$. This fact
and Eq. (\ref{diff}) imply that in the leading approximation
one can use the 2SC bar-quantities
in calculating $V_\Delta$ in the reduced potential. 
In other words,  
the effective potential can be decomposed into the ``constant'' 
2SC part $V_\Delta$, with frozen fermion parameters, and the dynamical 
gluonic part:
\begin{equation}
  \tilde{V}_{\rm eff} \to
  \tilde{V}_{\rm eff}(\bar{\Delta},\bar{\mu}_e,\bar{\mu}_8;B,C,D) = 
  V_\Delta(\bar{\Delta},\bar{\mu}_e,\bar{\mu}_8) + \frac{1}{2}M_B^2 B^2
  + T_{DB} D B^2 + \frac{1}{2}\lambda_{BC} B^2 C^2
  + \frac{1}{2}\lambda_{CD} C^2 D^2 .
  \label{V_min1}
\end{equation}
Eq.~(\ref{appr}) is the exact solution for the potential (\ref{V_min1}) 
and the energy density at the stationary point is 
\begin{equation}
  \tilde{V}_{\rm eff}(\bar{\Delta},\bar{\mu}_e,\bar{\mu}_8;
             B_{\rm sol},C_{\rm sol},D_{\rm sol})
 =V_\Delta + \frac{1}{6}M_B^2 B^2_{\rm sol} = 
  V_\Delta - \frac{(-M_B^2)^3}{54 T_{DB}^2}
  \left(\,-\frac{\lambda_{CD}}{\lambda_{BC}}\,\right) < V_\Delta .
\end{equation}
Therefore the gluonic vacuum is more stable than the 2SC one.

In the description of the dynamics with
vector condensates, there is a subtlety connected with the derivation of
a {\it physical} effective potential, whose minima correspond to
stable or metastable vacua. The point is that
although the gauge symmetry is gone in the unitary gauge,
the present theory still has constraints. In fact, it is a system with
second-class constraints, similar to the theory of a free 
massive vector field $A_{\mu}$ described by the Proca Lagrangian
(for a thorough discussion of systems with second-class 
constraints, see Sec. 2.3 in book \cite{GT}).
In such theories, while the Lagrangian formalism can be used
without introducing a gauge, the physical Hamiltonian
is obtained by explicitly resolving the constraints.  

In our case,
this implies that 
to obtain the physical effective potential,
one has to impose the Gauss's law constraint 
on the conventional
effective potential $\tilde{V}_{\rm eff}$ (\ref{V_min1}).  
This constraint amounts to integrating out the time-like components
$A_0^{(a)}$. In the present approximation,
the latter can be done by using their
equations of motion, which are reduced to Eq. (\ref{gap-eq-D})
for $D = g A^{(3)}_{0}$ in our case. Omitting  
the suppressed $DB^2$-term in this equation, we get   
\begin{equation}
  T_{DB} B^2 + \lambda_{CD}C^2 D = 0.
\end{equation}
It leads to
the physical effective potential without the non-dynamical degree of
freedom $D$:
\begin{equation}
  \tilde{V}_{\rm eff}^{\rm phys} = V_\Delta + \frac{1}{2}M_B^2 B^2
  + \frac{1}{2}\lambda_{BC} B^2 C^2
  - \frac{T_{DB}^2 B^4}{2\lambda_{CD}C^2} .
  \label{V_Gauss}
\end{equation}
It is easy to show that solution (\ref{appr}) is
a minimum by analyzing the curvature of 
$\tilde{V}_{\rm eff}^{\rm phys}$. Note that because of the
constraint in Eq. (\ref{constraint}), this
potential is bounded from below.

In the next section, we will calculate $M_B^2$, 
$T_{DB}$, $\lambda_{BC}$,
and $\lambda_{CD}$. In particular, it will be shown 
that constraint (\ref{constraint}) is satisfied near the critical point.

\subsection{Dynamics in one-loop approximation}
\label{sec6}

In this subsection, we determine the GL effective potential (\ref{V_min1})
in one-loop approximation and derive the dispersion relations for
quarks in the gluonic phase. 
The 2SC $V_\Delta$ part of the potential is 
known~\cite{Huang:2003xd},
\begin{eqnarray}
  V_\Delta (\Delta,\mu_e,\mu_8) &=&
    \frac{\Delta^2}{4G_\Delta} - \frac{\mu_e^4}{12\pi^2}
  - \frac{\mu_{ub}^4}{12\pi^2} - \frac{\mu_{db}^4}{12\pi^2}
  - \frac{\tilde{\mu}^4}{3\pi^2}
    \nonumber \\ &&
  - \frac{\Delta^2}{\pi^2}\left[\,
     \tilde{\mu}^2-\frac{1}{4}\Delta^2\,\right]\ln \frac{4\Lambda^2}{\Delta^2}
  - \frac{\Delta^2}{\pi^2}\left[\,\Lambda^2 -
    2\tilde{\mu}^2+\frac{1}{8}\Delta^2\,\right] , \quad (\delta\mu < \Delta).
\label{V_2SC}
\end{eqnarray}
Here $\Lambda$ is the ultraviolet cutoff in the NJL model and
$\mu_{ub}$, $\mu_{db}$, and $\tilde{\mu}$ are given in 
Eqs.~(\ref{beta-eq}) and (\ref{tildemu}).
For clarity of the presentation, the  
bars in $\Delta$, $\mu_{e}$ and $\mu_{8}$ were omitted 
[${\cal O}(\tilde{\mu}^2/\Lambda^2)$ and 
${\cal O}(\Delta^2/\Lambda^2)$ and higher terms are neglected
in this expression].
\footnote{For realistic values $\Lambda = (1.5 - 2.0)\mu$ that we use,
while the contribution of
the ${\cal O}(\Delta^2/\Lambda^2)$ terms are parametrically
suppressed,
one can show that the contribution of the 
${\cal O}(\tilde{\mu}^2/\Lambda^2)$ terms is numerically suppressed.}
Note that the color and electrical charge neutrality conditions 
in the 2SC solution yield~\cite{Huang:2003xd}
\begin{equation}
  \delta \mu = \frac{3}{10}\mu - \frac{1}{5}\mu_8, 
  \label{2SC1}
\end{equation}
and
\begin{equation}
  (\tilde{\mu}^2+\delta\mu^2)\mu_8 =
   -\tilde{\mu}\Delta^2\left(\,\ln\frac{2\Lambda}{\Delta}-1\,\right)
   +\tilde{\mu}(\delta\mu^2+\mu_8^2)-\frac{1}{3}\mu_8^3 ,
  \label{2SC2}
\end{equation}
which is consistent with the result of Ref.~\cite{Gerhold:2003js},
$\mu_8 \sim {\cal O}(\Delta^2/\mu)$,
in the case of 
$\delta\mu=0$.
The size of the diquark gap
$\Delta$ is essentially determined by tuning the NJL
coupling constant $G_\Delta$ and cutoff $\Lambda$.

In Appendix \ref{B},
after straightforward but tedious 
calculations of relevant one-loop diagrams from the fermion
trace in Eq.~(\ref{V}), we find the following relations  
in the region $\delta\mu < \Delta$:
\begin{eqnarray}
 \lambda_{BC} &=& \frac{1}{80\pi^2}\frac{\tilde{\mu}^2}{\Delta^2}
             \left[\,-1+8\frac{\delta\mu^2}{\Delta^2}
             \left(\,1-\frac{\delta\mu^2}{\Delta^2}\,\right)\,\right],
 \label{bc}  \\[3mm]
 \lambda_{CD} &=&
  -\frac{1}{g^2}-\frac{1}{18\pi^2}\frac{\tilde{\mu}^2}{\Delta^2},
 \label{cd}  \\[3mm]
 T_{DB} &=& \frac{\mu_8}{2g^2} + 
\frac{\mu_8}{24\pi^2}\frac{\tilde{\mu}^2}{\Delta^2} 
       \left(\,-1+8\frac{\delta\mu^4}{\Delta^4}\,\right)
      +\frac{\tilde{\mu}}{48\pi^2}
       \left(\,-1+4\frac{\delta\mu^2}{\Delta^2}
                 +8\frac{\delta\mu^4}{\Delta^4}\,\right) .
 \label{T}
\end{eqnarray}
Here the tree contribution of gluons
\begin{equation}
  V_g \equiv -{\cal L}_g = -\frac{1}{2}
  F_{0j}^{(a)} F_{0j}^{(a)} =
  -\frac{1}{2g^2}\mu_8^2 B^2 + \frac{1}{2g^2}\mu_8 DB^2
  - \frac{1}{8g^2}B^2 D^2 - \frac{1}{2g^2}C^2 D^2 
\label{free}
\end{equation}
was also taken into account. As to the coefficient $M_{B}^2$,
its expression follows directly from Eqs. (\ref{m_M4}) and 
(\ref{free}):
\begin{equation}
  M_B^2 = 
 \frac{1}{g^2}\left(\,-\mu_8^2 + m_{M,4}^2\,\right)=
    - \frac{\mu_8^2}{g^2}
    + \frac{\tilde{\mu}^2}{6\pi^2}
      \left(\,1-\frac{2\delta\mu^2}{\Delta^2}\,\right).
  \label{M_B} 
\end{equation}

We see that the coefficient $\lambda_{CD}$ in (\ref{cd})
is definitely negative.
The parameter $M_B^2$, which is expressed through 
the Meissner mass (\ref{m_M4}), is negative when
\begin{equation}
  \delta \mu > \delta\mu_{\rm cr}, \qquad
  \delta\mu_{\rm cr} = \frac{\Delta}{\sqrt{2}}
   \sqrt{1-\frac{3\pi}{2\alpha_s}\frac{\mu_8^2}{\tilde{\mu}^2}},
  \qquad \alpha_s \equiv \frac{g^2}{4\pi} .
\label{cr}
\end{equation}

Relation (\ref{2SC1}) and Eq.~(\ref{tildemu}) yield
\begin{equation}
  \tilde{\mu}=\frac{9}{10}\mu+\frac{2}{5}\mu_8,
\label{3}
\end{equation}
and, at the critical point, we find from Eqs. 
(\ref{2SC1}), (\ref{2SC2}) and (\ref{cr}), (\ref{3}) 
that $\mu_{8}$ is approximately
\begin{equation}
  \mu_8^{(\rm sol)} = \frac{3-\ln\frac{200\Lambda^2}{9\mu^2}}{12+
  \frac{4}{9}\left(\,
     \ln\frac{200\Lambda^2}{9\mu^2}-2\,\right)}\,\mu\, .
\end{equation}
For realistic values $\Lambda = (1.5-2.0) \mu$ and 
$\alpha_s=0.75-1.0$, we obtain numerically 
\begin{equation}
  \frac{3\pi}{2\alpha_s}
\frac{\mu_{8}^2}{\tilde{\mu}^2}|_{\mu_8 = \mu_8^{(\rm sol)}}
= 0.03\mbox{--}0.1\;.
\end{equation}
This implies that the tree gluon contribution in Eq. (\ref{cr}) 
decreases 
the value of $\delta\mu_{\rm cr}$ by 1.5\%--5\% in comparison to
its value in the (non-gauged) NJL model. The smallness of this
correction is in accordance 
with the dominance of hard-dense-loop diagrams.

Let us now turn to the coefficient $\lambda_{BC}$ (\ref{bc}). 
At the critical point 
$\delta\mu = \delta\mu_{\rm cr}$, it is [henceforth we will
not show explicitly the superscript ${(\rm sol})$ in $\mu_8$]:
\begin{equation}
 \lambda_{BC} = \frac{1}{80\pi^2}\frac{\tilde{\mu}^2}{\Delta^2}
\left(\,1-\frac{9\pi^2}{2\alpha_s^2}\frac{\mu_8^4}{\tilde{\mu}^4}\,\right).
\label{bc1}
\end{equation}
Because the $\mu_8^4/\tilde{\mu}^4$-term is negligibly small,
we conclude that the coefficient $\lambda_{BC}$ is positive
near the critical point.
Thus, constraint (\ref{constraint}) is satisfied indeed. 

Utilizing Eqs.~(\ref{bc})--(\ref{T}) and (\ref{M_B})
in Eq.~(\ref{appr}),
one can obtain the solutions for $B$, $C$, and $D$ in the 
near-critical region.
Indeed, neglecting higher order terms in $\mu_8/\mu$ in 
Eqs.~(\ref{bc})--(\ref{T}) and (\ref{M_B}),
we get the approximate relations
\begin{equation}
  M_B^2 \simeq \frac{\tilde{\mu}^2}{6\pi^2}
         \left(\,1-\frac{\delta\mu^2}{\delta\mu_{\rm cr}^2}\,\right), \quad
  \lambda_{BC} \simeq \frac{9}{160\pi^2}, \quad
  \lambda_{CD}\simeq -\frac{1}{4\pi\alpha_s}-\frac{1}{4\pi^2}, \quad
  T_{DB} \simeq  \frac{\tilde{\mu}}{16\pi^2}
     +\frac{\mu_8}{16\pi^2}\left(\,3+\frac{2\pi}{\alpha_s}\,\right), 
\label{apprT}
\end{equation}
which lead us to the near-critical solution:
\begin{eqnarray}
B_{\rm sol} &=& 
        \frac{\delta\mu^2-\delta\mu_{\rm cr}^2}{\delta\mu_{\rm cr}^2}
        \frac{16\,\tilde{\mu}\,
              \sqrt{10\left(\,1+\frac{\pi}{\alpha_s}\,\right)}}
             {27\left[\,1+\frac{\mu_8}{\tilde{\mu}}
                 \left(\,3+\frac{2\pi}{\alpha_s}\,\right)\,\right]},
 \label{b} \\[3mm]
C_{\rm sol} &=& \frac{\sqrt{\delta\mu^2-\delta\mu_{\rm cr}^2}}
                    {\delta\mu_{\rm cr}}\,
              \frac{4\sqrt{5}\,\tilde{\mu}\,}{9},
 \label{c} \\[3mm]
D_{\rm sol} &=& 
        \frac{\delta\mu^2-\delta\mu_{\rm cr}^2}{\delta\mu_{\rm cr}^2}
        \frac{8\,\tilde{\mu}}
             {9\left[\,1+\frac{\mu_8}{\tilde{\mu}}
               \left(\,3+\frac{2\pi}{\alpha_s}\,\right)\,\right]}. 
 \label{d}
\end{eqnarray}

It is noticeable that this solution 
describes nonzero field strengths $F_{\mu\nu}^{(a)}$
which correspond to the presence
of
{\it non-abelian}
constant chromoelectric-like condensates in the ground state:
\begin{eqnarray}
\VEV{E_{3}^{(2)}} &=& \VEV{F_{03}^{(2)}} 
= \frac{1}{g}\,C_{\rm sol}D_{\rm sol}\,,\\
\VEV{E_{3}^{(7)}} &=& \VEV{F_{03}^{(7)}} 
= \frac{1}{2g}\,B_{\rm sol}
                             \left(\,2\mu_{8} - D_{\rm sol}\,\right).
\label{E}
\end{eqnarray}

We emphasize that while
an abelian constant electric field in different
media always leads to an instability,
\footnote {In metallic and superconducting
media, such an instability is classical in its origin.
In semiconductors and insulators, this instability is
manifested in an creation of electron-hole
pairs through a quantum tunneling process.}
non-abelian constant
chromoelectric fields do not in many cases: For a thorough discussion
of the stability problem 
for constant $SU(2)$ non-abelian  fields in theories with 
zero baryon density, see Ref. \cite{Brown:1979bv}.  
On a technical side, this difference is
connected with that while a vector potential corresponding to
a constant abelian electric field depends on spatial and/or time
coordinates,
a constant non-abelian chromoelectric field is expressed through
constant vector potentials, as takes place in our case, and
therefore momentum and energy are good quantum numbers in the latter.

In order to illustrate the stability issue in the gluonic phase,
let us consider the dispersion relations for quarks there.
Because the vacuum expectation values (\ref{b})-(\ref{d}) are small
near the critical point and because red and green quarks are gapped
in the 2SC phase,
the dispersion relations for gapless blue up and down quarks
are of the most interest. 
From Eq. (\ref{Sg-inv}) we find
that up to the first order in $B^2$ they are
\begin{eqnarray}
  p^{0}_{ub} &=& |\vec p|-\mu_{ub}
  +\frac{B_{\rm sol}^2}{4}
   \frac{1}{2|\vec p|+\mu_8+\frac{\Delta^2}{2\tilde{\mu}-\mu_8}}
  -\frac{B_{\rm sol}^2}{4}\frac{(p^3)^2}{\vec p{\;}^2}\left(
    \frac{1}{2|\vec p|+\mu_8+\frac{\Delta^2}{2\tilde{\mu}-\mu_8}}
  +\frac{2(|\vec p|-\tilde{\mu})+\mu_8}
        {\Delta^2-\mu_8^2-2\mu_8(|\vec p|-\tilde{\mu})}\,\right),
 \label{bu} \\[3mm]
  p^{0}_{db} &=& |\vec p|-\mu_{db}
  +\frac{B_{\rm sol}^2}{4}
   \frac{1}{2|\vec p|+\mu_8+\frac{\Delta^2}{2\tilde{\mu}-\mu_8}}
  -\frac{B_{\rm sol}^2}{4}\frac{(p^3)^2}{\vec p{\;}^2}\left(
    \frac{1}{2|\vec p|+\mu_8+\frac{\Delta^2}{2\tilde{\mu}-\mu_8}}
  +\frac{2(|\vec p|-\tilde{\mu})+\mu_8}
        {\Delta^2-\mu_8^2-2\mu_8(|\vec p|-\tilde{\mu})}\,\right) .
 \label{bd}
\end{eqnarray}
The $B^2$-terms in Eqs.(\ref{bu}) and (\ref{bd}) 
lead to non-spherical Fermi surfaces determined by the
following equations: 
\begin{eqnarray}
 |\vec p| &=& \mu_{ub}
  -\frac{B_{\rm sol}^2\sin^2\theta}{4}
   \frac{1}{2\mu_{ub}+\mu_8+\frac{\Delta^2}{2\mu_{ub}+\mu_e+\mu_8}}
  -\frac{B_{\rm sol}^2\cos^2\theta}{4}
   \frac{\mu_e+\mu_8}
        {\Delta^2+\mu_8 \mu_e+\mu_8^2} , 
 \qquad \mbox{(blue up)}
 \label{fs-up} \\[3mm]
 |\vec p| &=& \mu_{db}
  -\frac{B_{\rm sol}^2\sin^2\theta}{4}
   \frac{1}{2\mu_{db}+\mu_8+\frac{\Delta^2}{2\mu_{db}-\mu_e+\mu_8}}
  +\frac{B_{\rm sol}^2\cos^2\theta}{4}
   \frac{\mu_e-\mu_8}
        {\Delta^2-\mu_8 \mu_e+\mu_8^2} , 
 \qquad \mbox{(blue down)}
 \label{fs-down}
\end{eqnarray}
where we neglected higher order terms in $B^2$ and 
defined the angle $\theta$,
\begin{equation}
  p^3 \equiv |\vec p|\cos\theta .
\end{equation}
The dispersion relations (\ref{bu}) and (\ref{bd}) clearly show that 
there is no instability in the quark sector in this problem.

As to bosonic degrees of freedom (gluons and composite bosons),
because it is very involved to derive their derivative terms
from the fermion loop in the gluonic phase,
this issue is beyond the scope of this paper. It is however noticeable
that there are no
instabilities for bosons in a phase with vector condensates in
the gauged $\sigma$-model with a chemical potential for hypercharge
\cite{sigmamodel}. Although that model is much simpler than
the present one, its phase with vector condensates has many common
features with the gluonic phase and this fact is encouraging.
Note that among light collective excitations in 
the gluonic phase, there should
be Nambu-Goldstone bosons connected with the spontaneous
breakdown of the $SO(3)_{\rm rot}$ and the $\tilde{\tilde{U}}(1)_{\rm em}$
(in the presence of photon field, the latter will be absorbed
in the electric (longitudinal) part of the field).

We emphasize that these constant color condensates in the gluonic
phase do {\it not} produce long range color {\it forces} acting on
quasiparticles.
This can be seen from the dispersion relations (\ref{bu}) and
(\ref{bd}) for quarks
in this model. They show that momentum and energy are conserved numbers.
It would be of course impossible in the presence of long range forces.
The role of these condensates is actually more dramatic: They change
the structure of the ground state, making it anisotropic and
(electrically) superconducting. Only in this sense, one can speak about a
long range character of the condensates.

\subsection{Searching for other solutions}
\label{sec7}

Are there solutions of Eqs. (\ref{gap-eq-B})-(\ref{gap-eq-D}) other 
than that found in the previous section and the trivial one with
$B_{\rm sol}=C_{\rm sol} =D_{\rm sol} =0$? We will address this question 
in this subsection.

First of all, it is easy to see that 
there are two {\it formally} nontrivial solutions with $B_{\rm sol}=0$:
a) $C_{\rm sol} = 0$, 
$D_{\rm sol}$-arbitrary and b) $D_{\rm sol} = 0$, $C_{\rm sol}$-arbitrary.
However, as follows from Eq. (\ref{BCD}), both these solutions
lead to zero $SU(2)_c$ field strength
$f_{\mu\nu}^{(l)}$ (\ref{su2strength}) and therefore they are 
gauge equivalent to the 2SC solution without vector condensates.

It is also easy to check that for $D_{\rm sol} = 0$,
Eqs. (\ref{gap-eq-B})-(\ref{gap-eq-D}) lead to 
$B_{\rm sol} = 0, C_{\rm sol} = 0$. Therefore
we conclude that both $B$ and $D$ have to be nonzero in a nontrivial
solution. Physics underlying this conclusion is clear.
When $B_{\rm sol}\equiv g\VEV{A_z^{(6)}} \ne 0$, 
the color charge densities
of red and green quarks are generally different, so that
$D_{\rm sol} \equiv g\VEV{A_0^{(3)}} \ne 0$ is also required
for color neutrality
(recall that $gA_0^{(3)}$ can be considered as a chemical
potential $\mu_3$ related to the third component of
the color isospin). 

There does exist one additional nontrivial solution
with nonzero $\VEV{E_{3}^{(7)}} = 
\VEV{F_{03}^{(7)}}$ (see Eq. (\ref{E})): 
\begin{equation}
  C_{\rm sol}=0, \quad 
D_{\rm sol}=-\frac{T_{DB}}{\lambda_{BD}}, \quad
B^2_{\rm sol} = -\frac{M_B^2}{\lambda_B} 
+ \frac{T_{DB}^2}{\lambda_B\lambda_{BD}}.
\label{new}
\end{equation}
This solution corresponds to a phase in which while
the gauge $SU(2)_c$  symmetry and 
the rotational $SO(3)_{\rm rot}$ are broken,
the electromagnetic $\tilde{\tilde{U}}(1)_{\rm em}$ is exact.
While physics in this phase is quite interesting, there is
the following problem in justifying
its existence. As follows from Eq. (\ref{apprT}),
the coefficient $T_{DB}$ does not approach zero
at the critical point $M_B^2 = 0$. This and
Eq. (\ref{new}) imply that the values of $B_{\rm sol}$ and 
$D_{\rm sol}$ in this solution are also nonzero at the
critical point. Therefore the solution 
corresponds to the first order phase transition and
the GL approach is not appropriate in this case.
Since the derivation of Eqs. (\ref{gap-eq-B})-(\ref{gap-eq-D}),
(\ref{V_min1}), and (\ref{V_Gauss})
was based on this approach, all these equations themselves
will be modified in this phase. Therefore 
the question concerning the existence of 
solution (\ref{new}) is open.
It seems that a numerical analysis would be the only  
reliable way to answer it.

In conclusion, the following two remarks are in order.
a) Since both solutions (\ref{b})-(\ref{d}) and 
(\ref{new}) 
are cylindrically symmetric, it will
be appropriate to call the phase corresponding
to solution (\ref{b})-(\ref{d}) as
a {\it gluonic cylindrical phase} I 
and the phase corresponding to solution (\ref{new})
as a {\it gluonic cylindrical phase} II.
b) The fact that any nontrivial solution should
have both $B$ and $D$ to be nonzero follows 
from the presence of the triple vertex $T_{DB}DB^2$
in the GL potential: for $B \neq 0$, this vertex
inevitably leads to a nonzero $D$. One can call this 
a tadpole mechanism: For a given $B$, the diagram
corresponding to such a vertex is a tadpole with
a fermion loop (with $B$ insertions), 
producing the coefficient $T_{DB}$,
and with a tail being the field $D$.

\subsection{Confinement picture and exotic hadrons in gluonic phase}
\label{sec8}

In this subsection, we will describe some additional features of
the gluonic phase. In particular, we will describe the confinement
picture, which can be appropriate for the description of its dynamics in 
the infrared region,
with the energy scale
of order  $\Lambda'_{\rm QCD} \lesssim {\cal O}(\mbox{10 MeV})$
(or even much smaller) \cite{Rischke:2000cn}.
As we will see, the dynamics 
in this dense medium includes light {\it exotic} vector mesons
some of which can condense.

The
gluon condensates 
are mostly generated at energy scales between the
confinement scale in the 2SC state  $\Lambda'_{\rm QCD}$
and the quark chemical potential, 
which is about
300-500 MeV. It is the same region where the chromomagnetic instability in
the 2SC phase is created and where the hard dense loop approximation
is (at least qualitatively) reliable. 
At such scales, gluons are still
appropriate dynamical degrees of freedom and utilizing the Higgs approach
with color condensates in a particular gauge is appropriate and
consistent: It is a region of hard physics. Because the 
gluonic cylindrical phase I 
occurs as a result of a conventional second order phase transition, the
gluon condensates are very small only in the immediate 
surroundings of the critical point 
$\delta\mu = \bar{\Delta}/\sqrt{2}$. 
Outside that region, their values should be of the order of the
typical scale $\delta\mu \sim {\bar\Delta}\sim 50-100$ MeV. As to the
gluonic cylindrical phase II, because it is connected with 
a first order phase transition, the situation depends on whether
it is a strong or a weak one. While for the former,
one could expect that the
gluon condensates are of order  $\delta\mu \sim {\bar\Delta}$ even at the
nearest surrounding of the critical point [see Eq. (\ref{new})],
for the latter, they could be of order $\Lambda_{QCD}'$ there,
modulo the question of the existence of this phase (see the discussion in
the previous subsection).

The description of the dynamics in the gluonic phases in the infrared region
depends on the
value of the gluon condensates. If they are essentially larger than
$\Lambda'_{\rm QCD}$,
then the
Higgs description is appropriate even in the infrared region,
similarly as it happens in the electroweak theory. 
Indeed, 
when the $SU(2)_c$ gauge symmetry becomes completely broken
by the dynamics with a characteristic scale being
essentially larger than 
$\Lambda'_{\rm QCD}$,
the strong coupling dynamics presented
in the 2SC solution at the scale of order $\Lambda'_{\rm QCD}$
is washed out.
In this regard, the gluonic phases with large vector condensates
are similar to the color-flavor locked (CFL) phase
with a large quark chemical potential $\mu$, where the 
color condensates (although not vector ones) completely 
break the $SU(3)_c$ color gauge symmetry \cite{Alford:1998mk}. 

But what happens if the gluon condensates are small, 
$\alt \Lambda'_{\rm QCD}$? This regime corresponds to
the nearcritical dynamics in the gluonic cylindrical phase I
and the confinement picture should be appropriate
for the description of the infrared dynamics in this case.

In order to answer this question, note the following.
As one can see in Tables \ref{tab1} and \ref{tab2},
the electric charge
$\tilde{\tilde{Q}}_{\rm em}$ and the
baryon number $\tilde{\tilde{\cal B}}$ 
are integer both for gluons and quarks.
Do they describe hadronic-like excitations? 
We believe that the answer is ``yes". 
The point is that in models with Higgs fields
in the (anti-) fundamental representation of the gauge group, there is no
phase transition between Higgs and confinement phases
\cite{Osterwalder:1977pc,Fradkin:1978dv,Dimopoulos:1980hn}
and this is the case in the present model. Indeed,
in the 2SC phase, the breakdown  
$SU(3)_c \to SU(2)_c$ is triggered by the diquark condensate,
which is assigned to the anti-fundamental representation of the $SU(3)_c$,
and gauge $SU(2)_c$ symmetry breaking occurs
when the $SU(2)_c$ doublet vector field $\phi_\mu$ develops the VEV.
Because of that, we can apply the complementarity principle 
\cite{Osterwalder:1977pc,Fradkin:1978dv,Dimopoulos:1980hn} for
the description of the dynamics in the gluonic phase with 
small condensates in the infrared region. 
What matters is the existence
of the unitary gauge given by constraints (\ref{2SC}) and
(\ref{unitary}). In this gauge all gauge dependent degrees of freedom
are removed. 

Due to the complementarity principle, the Higgs and confinement phases
provide dual, and physically equivalent, descriptions of 
dynamics. In particular, they
provide two complementary descriptions of a spontaneous breakdown of 
global symmetries, such as the rotational $SO(3)$ 
and the electromagnetic $U(1)$ in the gluonic cylindrical phase I
and the rotational $SO(3)$ in the gluonic cylindrical phase II.
Following Ref. \cite{Dimopoulos:1980hn}, we will consider
the dual, gauge invariant,
approach in this model and show that all the gluonic and quark fields 
can indeed be replaced by colorless composite ones. 

The flavor quantum numbers of these composite fields
are described by the conventional 
electric and baryon charges $Q_{\rm em}$ and ${\cal B}$. They are 
integer and coincide
with those the operators $\tilde{\tilde{Q}}_{\rm em}$ and 
${\tilde{\tilde{\cal B}}}$ yield for gluonic and quark fields. 
The composite fields in confinement picture should coincide with
the corresponding fields in the Higgs picture in the unitary gauge
in the classical approximation.

The nonzero VEVs in the Higgs picture (common for these two
gluonic phases) are: 
\footnote{Recall that we do not show explicitly the superscript
$({\rm sol})$ in the chemical potentials $\mu_3$ and  $\mu_8$.}
\begin{equation}
  \VEV{\bm{\Phi}}=(0,0,\bar{\Delta})^T, \quad
  \mu_3 \equiv D_{\rm sol}=g\VEV{A_0^{(3)}}, \quad
  \mu_8 \equiv \frac{\sqrt{3}}{2} g\VEV{A_0^{(8)}}, \quad
   B_{\rm sol}=g\VEV{A_z^{(6)}},
\end{equation}
and thereby\footnote{Recall that the diquark field $\bm{\Phi}$
is an anti-triplet under the $SU(3)_c$ symmetry.}
\begin{equation}
  \VEV{iD_z \bm{\Phi}^*}=(0,B_{\rm sol}\bar{\Delta}/2,0)^T, \quad
   g\VEV{F_{0z}^{7}} = 
 B_{\rm sol}\left(\mu_8 -
 \frac{\mu_3}{2}\right), \quad
   g\VEV{F_{0z}\bm{\Phi}^*} =
   -i\frac{B_{\rm sol}\bar{\Delta}}{4}
\left(2\mu_8 -\mu_3 \right)(0,1,0)^T,
\end{equation}
where
\begin{equation}
  \bm{\Phi} \equiv
   \left(
    \begin{array}{@{}c@{}}
      \Phi^r \\ \Phi^g \\ \Phi^b
    \end{array}
  \right) , \quad
  F_{\mu\nu} \equiv F_{\mu\nu}^{(a)} T^a,  \quad T^a \equiv \frac{\lambda^a}{2}
\end{equation}
with $\lambda^a$'s being the $SU(3)_c$ Gell-Mann matrices.
In the classical approximation,
we replace the above fields by their VEVs.
Then the following composite fields can be written in terms of
the elementary fields:
\begin{equation}
  D_\mu \bm{\Phi}^* \to -\frac{ig}{2}\bar{\Delta}
   \left(
    \begin{array}{@{}c@{}}
      A_\mu^{(4)} - iA_\mu^{(5)} \\[3mm] A_\mu^{(6)} - iA_\mu^{(7)} \\[3mm]
     {\displaystyle -\frac{2}{\sqrt{3}}A_\mu^{(8)}}
    \end{array}
  \right) ,
\end{equation}
\begin{equation}
  F_{\mu z} \bm{\Phi}^* \to -\frac{i}{4}B_{\rm sol}\bar{\Delta}
   \left(
    \begin{array}{@{}c@{}}
      A_\mu^{(1)} - iA_\mu^{(2)} \\[3mm]
     -A_\mu^{(3)} + \sqrt{3} A_\mu^{(8)} \\[3mm]
     2i A_\mu^{(7)}
    \end{array}
  \right) ,
\end{equation}
and
\begin{equation}
  F_{0j} \bm{\Phi}^* \to -\frac{i}{4}{\bar\Delta}
   \left(
    \begin{array}{@{}c@{}}
      (2\mu_8 +\mu_3)
(A_j^{(4)} - iA_j^{(5)}) \\[3mm]
   (2\mu_8 -\mu_3)(A_j^{(6)} - iA_j^{(7)}) \\[3mm]
       0
    \end{array}
  \right) .
\end{equation}

By using the above relations,
we can construct composite fermions and bosons in confining picture.
For example,
the (up and down) blue quark fields can be rewritten
as $\bm{\Phi}^T \bm{\psi}_i$
with $\bm{\psi}_{i} \equiv (\psi_{ir},\psi_{ig},\psi_{ib})^T$,
$(i=u,d)$.
Note that in the classical approximation
the composite fields $\bm{\Phi}^T \bm{\psi}_i$ yield
$\bar{\Delta} \psi_{ib}$.
The green quarks in the confinement picture can be described by
$(D_z \bm{\Phi}^*)^\dagger \bm{\psi}_i \to 
\frac{1}{2}B_{\rm sol}\bar{\Delta} \psi_{ig}$.
By using the epsilon tensor, we can rewrite the red quarks as
$\epsilon^{\alpha\beta\gamma} \psi_{i\alpha} (D_z \bm{\Phi}^*)_\beta
 (\bm{\Phi}^*)_\gamma \to 
\frac{1}{2}B_{\rm sol}\bar{\Delta}^2 \psi_{ir}$.
For the diquark field, only the real part of the anti-blue one is
a physical degree of freedom.
The composite field is
$\bm{\Phi}^\dagger \bm{\Phi} \to 2\bar{\Delta} \mbox{Re}\Phi^b$.
For other fields, see Table \ref{tab3}.

Similarly, we can construct the vector composite fields.
For example, we find
$(D_z \bm{\Phi}^*)^\dagger (D_\mu \bm{\Phi}^*) \to
 \frac{1}{4}
B_{\rm sol} \bar{\Delta}^2 (A_\mu^{(6)}-iA_\mu^{(7)}) \sim \phi_\mu^g$,
$\epsilon^{\alpha\beta\gamma} 
(D_\mu \bm{\Phi}^*)_\alpha (D_z \bm{\Phi}^*)_\beta
 (\bm{\Phi}^*)_\gamma \to \frac{1}{4}
B_{\rm sol}\bar{\Delta}^3 (A_\mu^{(4)}-iA_\mu^{(5)}) \sim
 \phi_\mu^r$,
$\epsilon^{\alpha\beta\gamma} (F_{z\mu}\bm{\Phi}^*)_\alpha
  (D_z\bm{\Phi}^*)_\beta (\bm{\Phi}^*)_\gamma \to
  \frac{1}{8}
B^2_{\rm sol}\bar{\Delta}^3 (A_\mu^{(1)}-iA_\mu^{(2)}) \sim {\cal A}_\mu^-$,
etc.. We summarize them in Table \ref{tab4}.

Some of vector mesons in Table \ref{tab4} are exotic because
they carry baryon charge.
(In vacuum QCD, mesons carry of course no baryon charge.)
For example, the electric and baryon charges $Q_{\rm em}$ 
and ${\cal B}$ of vector mesons corresponding to
$A^{(\pm)}_{\mu} = A^{(1)}_{\mu} \pm iA^{(2)}_{\mu}$ gluons
are equal to $\pm 1$ and $\pm 2$, respectively. The origin
of these exotic quantum numbers is connected with (anti-)
diquarks,
which are constituents of these mesons (see Table \ref{tab4}).
Indeed, (anti-) diquarks are bosons carrying the baryon charge
$\pm 2/3$ and therefore are exotic themselves.

This feature has a dramatic consequence for the gluonic cylindrical
phase I. Since in the Higgs description of 
this phase $A^{(\pm)}$ gluons are condensed (leading to 
the spontaneous $\tilde{\tilde{U}}(1)_{\rm em}$ breakdown),
we conclude that in the confinement picture
this corresponds to a condensation of {\it exotic} vector mesons.
In this regard,
it is appropriate to mention that some authors speculated about
a possibility of a condensation of vector $\rho$ mesons in dense
baryon matter \cite{rho}. 
The dynamics in the gluonic phase 
yields a scenario even with a more unexpected condensation.

\begin{table}
  \begin{tabular}{|c|cc|c|}\hline
          & $Q_{\rm em}$ & ${\cal B}$ & $I_3$ \\
    \hline \hline
     $\epsilon^{\alpha\beta\gamma} \psi_{u\alpha} (D_z\bm{\Phi}^*)_\beta
       (\bm{\Phi}^*)_\gamma \sim \psi_{ur}$
           & $0$ & $-1$ & $\frac{1}{2}$ \\
     $(D_z\bm{\Phi}^*)^\dagger \bm{\psi}_u \sim \psi_{ug}$
           & $1$ & $1$ & $\frac{1}{2}$ \\ \hline\hline
    $\epsilon^{\alpha\beta\gamma} \psi_{u\alpha} (F_{0z}\bm{\Phi}^*)_\beta
      (\bm{\Phi}^*)_\gamma \sim \psi_{ur}$
          & $0$ & $-1$ & $\frac{1}{2}$ \\
    $(F_{0z}\bm{\Phi}^*)^\dagger \bm{\psi}_{u} \sim \psi_{ug}$
          & $1$ & $1$ & $\frac{1}{2}$ \\ \hline\hline
    $\bm{\Phi}^T \bm{\psi}_u \sim \psi_{ub}$
          & $1$ & $1$ & $\frac{1}{2}$ \\ \hline\hline
    $\epsilon^{\alpha\beta\gamma} \psi_{d\alpha} (D_z\bm{\Phi}^*)_\beta
      (\bm{\Phi}^*)_\gamma \sim \psi_{dr}$
          & $-1$ & $-1$ & $-\frac{1}{2}$ \\
    $(D_z\bm{\Phi}^*)^\dagger \bm{\psi}_d \sim \psi_{dg}$
          & $0$ & $1$ & $-\frac{1}{2}$ \\ \hline\hline
    $\epsilon^{\alpha\beta\gamma} \psi_{d\alpha} (F_{0z}\bm{\Phi}^*)_\beta
      (\bm{\Phi}^*)_\gamma \sim \psi_{dr}$
          & $-1$ & $-1$ & $-\frac{1}{2}$ \\
    $(F_{0z}\bm{\Phi}^*)^\dagger \bm{\psi}_d \sim \psi_{dg}$
          & $0$ & $1$ & $-\frac{1}{2}$ \\ \hline\hline
    $\bm{\Phi}^T \bm{\psi}_d \sim \psi_{db}$
          & $0$ & $1$ & $-\frac{1}{2}$ \\ \hline\hline
    $\bm{\Phi}^\dagger \bm{\Phi} \sim \mbox{Re}\Phi^b$
          & $0$ & $0$ & $0$ \\ \hline\hline
    $(F_{0z}\bm{\Phi}^*)^\dagger (F_{0z}\bm{\Phi}^*) \sim \mbox{Re}\Phi^b$
          & $0$ & $0$ & $0$ \\ \hline
  \end{tabular}
  \caption{Composite fermions and scalar fields in
           the confinement picture. \label{tab3}}
\end{table}
\begin{table}
  \begin{tabular}{|c|cc|c|}\hline
          & $Q_{\rm em}$ & ${\cal B}$ & $I_3$ \\ \hline \hline
    $\epsilon^{\alpha\beta\gamma} (F_{z\mu}^*\bm{\Phi})_\alpha
      (D_z^*\bm{\Phi})_\beta (\bm{\Phi})_\gamma \sim {\cal A}_\mu^+$
    & $1$ & $2$ & 0 \\
    $(D_z\bm{\Phi}^*)^\dagger (F_{z\mu}\bm{\Phi}^*) + (\mbox{h.c.})
     \sim A_\mu^{(3)}, A_\mu^{(8)}$
    & 0 & 0 & 0 \\
    $\epsilon^{\alpha\beta\gamma} (F_{z\mu}\bm{\Phi}^*)_\alpha
      (D_z\bm{\Phi}^*)_\beta (\bm{\Phi}^*)_\gamma \sim {\cal A}_\mu^-$
    & $-1$ & $-2$ & 0 \\  \hline\hline
    $\epsilon^{\alpha\beta\gamma} (F_{0j}^* D_z^*\bm{\Phi})_\alpha
      (D_z^*\bm{\Phi})_\beta (\bm{\Phi})_\gamma \sim {\cal A}_j^+$
    & $1$ & $2$ & 0 \\
    $i(D_z\bm{\Phi}^*)^\dagger (D_\mu D_z\bm{\Phi}^*) + (\mbox{h.c.})
     \sim A_\mu^{(3)}, A_\mu^{(8)}$
    & 0 & 0 & 0 \\
    $\epsilon^{\alpha\beta\gamma} (F_{0j}D_z\bm{\Phi}^*)_\alpha
      (D_z\bm{\Phi}^*)_\beta (\bm{\Phi}^*)_\gamma \sim {\cal A}_j^-$
    & $-1$ & $-2$ & 0 \\  \hline\hline
    $\epsilon^{\alpha\beta\gamma} (D_\mu^*\bm{\Phi})_\alpha
      (D_z^*\bm{\Phi})_\beta \bm{\Phi}_\gamma \sim \phi_\mu^{*r}$
    & $1$ & $2$ & 0 \\
    $(D_z^*\bm{\Phi})^\dagger (D_\mu^*\bm{\Phi}) \sim \phi_\mu^{*g}$
     & $0$ & $0$ & 0 \\ \hline\hline
    $\epsilon^{\alpha\beta\gamma} (F_{0j}^*\bm{\Phi})_\alpha
      (D_z^*\bm{\Phi})_\beta (\bm{\Phi})_\gamma \sim \phi_j^{*r}$
    & $1$ & $2$ & 0 \\
    $(D_z^*\bm{\Phi})^\dagger (F_{0j}^*\bm{\Phi}) \sim \phi_j^{*g}$
    & $0$ & $0$ & 0 \\ \hline\hline
    $\epsilon^{\alpha\beta\gamma} (D_\mu\bm{\Phi}^*)_\alpha
      (D_z\bm{\Phi}^*)_\beta (\bm{\Phi}^*)_\gamma \sim \phi_\mu^r$
    & $-1$ & $-2$ & 0 \\
    $(D_z\bm{\Phi}^*)^\dagger (D_\mu\bm{\Phi}^*) \sim \phi_\mu^g$
    & $0$ & $0$ & 0 \\ \hline\hline
    $\epsilon^{\alpha\beta\gamma} (F_{0j}\bm{\Phi}^*)_\alpha
      (D_z\bm{\Phi}^*)_\beta (\bm{\Phi}^*)_\gamma \sim \phi_j^r$
    & $-1$ & $-2$ & 0 \\
    $(D_z\bm{\Phi}^*)^\dagger (F_{0j}\bm{\Phi}^*) \sim \phi_j^g$
    & $0$ & $0$ & 0 \\ \hline\hline
    $\bm{\Phi}^T (iD_\mu \bm{\Phi}^*) + \mbox{(h.c.)} \sim
      A_\mu^{(8)}$ & $0$ & $0$ & $0$ \\ \hline
  \end{tabular}
  \caption{Composite vectors in the confinement picture.
           \label{tab4}}
\end{table}

\section{Dynamics with gluon condensates: General analysis and
classification}
\label{general}

In the previous sections, the dynamics with gluon condensates 
connected with the instability for the
4-7th gluons in the 2SC phase was considered. The question
is whether there are gluonic phases other than those two
described in Sec. \ref{gluon}. 
In particular, an interesting issue is 
the dynamics of gluon condensates connected with
the instability for the 8th gluon in the g2SC phase. Moreover,
our consideration of the gluonic phase in Sec. \ref{gluon}
was somewhat heuristic. We used ansatz (\ref{BCD}) without
addressing the question whether it is self consistent, i.e.,
whether for VEVs $B, C, D \neq 0$ the equation of motions will 
or will not
lead to nonzero VEVs of other gluon fields. 
These questions will be addressed in the present section.
The general analysis we will use is based on the symmetry
consideration and the GL effective theory approach.

\subsection{Symmetry breaking structure}
\label{sec9}

In this subsection, we consider possible symmetry breaking samples
in the gauged NJL model and the structure of the corresponding homogeneous
gluon condensates. Our strategy is the following.
For each symmetry breaking sample, we will
pick up the maximal set of gluon condensates consistent with it.
The dynamics of course could allow a subset of this set to be
a solution. This much harder issue is intimately connected with
the structure of the GL effective theory corresponding
to the symmetry sample and will be discussed in 
Subsec. \ref{sec10}.

In the present analysis, it will be convenient to use the 
(partly-unitary) gauge (\ref{2SC}) for the diquark
field $\Phi^{\alpha}$.
Then the symmetry in the 2SC phase is that presented
in Eq. (\ref{sym-2SC}).
In the general case, the homogeneous gluon condensates consist of 32 VEVs,
$\VEV{A_\mu^{(a)}}$, $(a=1,2,\cdots,8, \mu=0,x,y,z)$.
The symmetry (\ref{sym-2SC}) contains 9 parameters. However,
since all gluon fields are singlet with respect to the 
chiral group $[U(1)_{\tau^3_L} \times U(1)_{\tau^3_R}]_{\chi}$,
only 7 parameters connected with the group
$SU(2)_c \times \tilde{U}(1)_{\rm em} \times SO(3)_{\rm rot}$
matter. By using the corresponding transformations,
the 32 VEVs can be reduced to 25 ones.

Let us show that these 25 VEVs can be chosen in the form: 
\begin{subequations}
\label{VEV-g}
\begin{equation}
 \VEV{A_z^{(8)}}\, ,
\label{seta}
\end{equation}
\begin{equation}
  \VEV{\phi_{x}} \; = \;
  \frac{1}{\sqrt{2}}
  \left(
  \begin{array}{@{}c@{}}
  \VEV{A_x^{(4)}} - i \VEV{A_x^{(5)}} \\[2mm]
  \VEV{A_x^{(6)}} - i \VEV{A_x^{(7)}}
  \end{array}
  \right) , \quad
  \VEV{\phi_{y}} \; = \;
  \frac{1}{\sqrt{2}}
  \left(
  \begin{array}{@{}c@{}}
  \VEV{A_y^{(4)}} \\[2mm] \VEV{A_y^{(6)}}
  \end{array}
  \right) , \quad
  \VEV{\phi_{z}} \; = \;
  \frac{1}{\sqrt{2}}
  \left(
  \begin{array}{@{}c@{}}
  0 \\[2mm] \VEV{A_z^{6}}
  \end{array}
  \right) ,
\end{equation}
\begin{equation}
  \VEV{A_j^{(\ell)}}, \quad (\ell=1,2,3, \; j=x,y,z),
\label{setb}
\end{equation}
and
\begin{equation}
  \VEV{A_0^{(a)}}  \quad (a=1,2,\cdots,8),
\label{setc}
\end{equation}
\end{subequations}
while 
$\VEV{A_x^{(8)}}=\VEV{A_y^{(8)}}=\VEV{A_y^{(5)}}=\VEV{A_y^{(7)}}=
\VEV{A_z^{(4)}}=\VEV{A_z^{(5)}}=\VEV{A_z^{(7)}}=0$.

The proof is going as follows. Because of the rotational symmetry,
without loss of generality, we can choose one direction for
the VEV of the spatial component of the 8th gluon, say,
$\VEV{A_z^{(8)}} \ne 0$.
We may apply the $SU(2)_c$ symmetry to the same spatial component of
$\phi_j$ and thereby obtain $\VEV{\phi_{z}^T} \sim (0\, ,\VEV{A_z^{(6)}})$.
Note that we can still rotate $\phi_x$ and $\phi_y$ by
a $SO(2)$
spatial rotation around $z$-axis and vary
the upper component of $\phi_j$
by a $U(1)$ transformation with the generator which is
an appropriate linear combination of $\sigma^3$ 
in the color $SU(2)_c$ 
and the charge $\tilde{Q}_{\rm em}$.
By using these transformations,
we can choose $\VEV{\phi_y}$ to be real.
As a result, we can reduce general homogeneous gluon condensates to
the 25 VEVs indicated above.

This set of 25 VEVs breaks symmetry (\ref{sym-2SC}) down to the
chiral $[U(1)_{\tau^3_L} \times U(1)_{\tau^3_R}]_{\chi}$
[henceforth this irrelevant for our discussion chiral 
group will be omitted].
When we take subsets of the set in Eq.~(\ref{VEV-g}), 
typical symmetry breaking patterns are: 
\footnote{
It is mathematically possible to consider also
the phase with $\VEV{A_0^{(3)}} \ne 0, \VEV{A_0^{(8)}} \ne 0$
and $\VEV{A_z^{(3)}} \ne 0, \VEV{A_z^{(8)}} \ne 0$,
in which the unbroken subgroup is
$U(1)_c \times \tilde{U}(1)_{\rm em} \times SO(2)_{\rm rot}$
with the generator $\sigma^3$ for the color $U(1)_c$.
However, the Debye and Meissner screening masses for the $SU(2)_c$ gluons
$A_\mu^{1,2,3}$ do not take imaginary values both
for $\delta \mu < \bar{\Delta}$ and $\delta \mu > \bar{\Delta}$.
Therefore there is no reason to consider such a phase.}
\begin{align*}
  SU(2)_c \times \tilde{U}(1)_{\rm em} \times SO(3)_{\rm rot} && \\
  & \to SU(2)_c \times \tilde{U}(1)_{\rm em} \times SO(2)_{\rm rot},
  & \mbox{(phase A),} \\[2mm]
  & \to \tilde{\tilde{U}}(1)_{\rm em} \times SO(2)_{\rm rot},
  & \mbox{(phase B),} \\[3mm]
  & \to SO(2)_{\rm rot}, & \mbox{(phase C),} \\[3mm]
  & \to SO(2)_{\rm diag}, & \mbox{(phase D),} \\[3mm]
  & \to \tilde{\tilde{U}}(1)_{\rm em}, & \mbox{(phase E),} \\[3mm]
  & \to \mbox{nothing}, & \mbox{(phase F).}
\end{align*}
The maximal subsets of the gluon condensates consistent with
these symmetry breaking patterns are
\begin{eqnarray*}
&&  \mbox{(phase A)} \quad
    \VEV{A_0^{(8)}} \ne 0, \quad \VEV{A_z^{(8)}} \ne 0, \\[3mm]
&&  \mbox{(phase B)} \quad
    \VEV{A_0^{(3)}} \ne 0, \VEV{A_0^{(6)}} \ne 0,
    \VEV{A_0^{(7)}} \ne 0, \VEV{A_0^{(8)}} \ne 0, \quad
    \VEV{A_z^{(3)}} \ne 0, \VEV{A_z^{(6)}} \ne 0,
    \VEV{A_z^{(8)}} \ne 0, \\[3mm]
&&  \mbox{(phase C)} \quad
    \VEV{A_0^{(a)}} \ne 0, (a=1,2,\cdots,8), \quad
    \VEV{A_z^{(1)}} \ne 0, \VEV{A_z^{(3)}} \ne 0,
    \VEV{A_z^{(6)}} \ne 0, \VEV{A_z^{(8)}} \ne 0, \\[3mm]
&&  \mbox{(phase D)} \quad
    \VEV{A_0^{(2)}} \ne 0, \VEV{A_0^{(8)}} \ne 0,  \quad
    \VEV{A_y^{(4)}} = \VEV{A_z^{(6)}} \ne 0, \\[3mm]
&&  \mbox{(phase E)} \quad
    \VEV{A_0^{(3)}} \ne 0, \VEV{A_0^{(6)}} \ne 0,
    \VEV{A_0^{(7)}} \ne 0, \VEV{A_0^{(8)}} \ne 0, \quad
    \VEV{A_j^{(3)}} \ne 0, \VEV{A_j^{(6)}} \ne 0, 
    \VEV{A_x^{(7)}} \ne 0, \VEV{A_z^{(8)}} \ne 0, \\[3mm]
&&  \mbox{(phase F)} \quad
\mbox{All 25 VEVs in Eqs. (11a)-(11d)} \ne 0
\end{eqnarray*}

As was pointed in \cite{Gorbar:2005rx,Gorbar:2005tx},
the phase A corresponds to the single plane-wave LOFF
phase \cite{Alford:2000ze}.
In this case, both color electric and magnetic field strengths 
equal zero.
This simplest case of the LOFF phase has been analyzed
by several authors~\cite{Alford:2000ze,Giannakis:2004pf,Gorbar:2005tx}.
With the neutrality conditions taken
into account, it was shown that
the single plane-wave LOFF phase cannot resolve the chromomagnetic
instability in the 2SC and g2SC regions~\cite{Gorbar:2005tx}.

While in the phase B the symmetry breakdown sample is the same
as in the gluonic cylindrical phase II considered in
Subsec. \ref{sec7}, in the phase C, it coincides with that
in the gluonic cylindrical phase I discussed in Sec.
\ref{gluon}. The VEVs of gluon fields in the 
latter constitute a subset of the VEVs presented
in the maximal set
in the phase C. Based on the GL approach, it
will be shown in Subsec. \ref{sec10} that 
this subset is self consistent. On the other hand, as was
already pointed out in Subsec. \ref{sec7}, the question
concerning the existence of the gluonic cylindrical phase II
is open. Recall that there are nonzero chromoelectric field strengths
in these phases.

 In the phase D, the VEV set does not change under
the  $SO(2)_{\rm diag}$ transformations with the generator
which is an appropriate linear combination of $\sigma^2$ in
the color  $SU(2)_c$ and the generator of spatial $y$-$z$ rotations.
We may call this phase the {\it gluonic color-spin locked (CSL) phase}.
In this phase, while nonzero $\VEV{A_0^{(2)}}$ and
$\VEV{A_0^{(8)}}$ are allowed, nonzero
$\VEV{A_0^{(1)}}$ and
$\VEV{A_0^{(3),\cdots,(7)}}$  are not, because they break
the $SO(2)_{\rm diag}$ symmetry.
It is noticeable that there exist both chromoelectric and
chromomagnetic field strengths in the gluonic CSL phase.

In the phase E, the $\tilde{\tilde{U}}(1)_{\rm em}$ is 
unbroken, while
the rotational $SO(3)_{\rm rot}$ symmetry is completely broken down.
At last, the phase F contains the maximum number, 25, of 
gluon condensates. This classification provides a useful
framework for studying dynamics of the gluon condensation.

In conclusion,
we comment on the relation between the homogeneous gluon condensates
and inhomogeneous diquark condensates. It is clear that when
there are no field strengths, constant gluon condensates can be
removed by using an appropriate gauge transformation. However,
the point
is that such gauge transformations can break constraint
(\ref{2SC}) for the diquark field. It is exactly what happens
in the case of the single plane-wave LOFF state (the phase A): One
can remove $\VEV{A_z^{(8)}}$ at the cost of introducing an
exponential factor 
with a linearly depending on $z$ phase in the diquark field $\Delta$.
For other examples, see Ref. \cite{Hashimoto:2006mn}.

\subsection{Ginzburg-Landau approach}
\label{sec10}

In this section, we study the GL approach
for dynamics with vector gluon condensates in the gauged NJL
model. As was shown in Sec. \ref{sec3}, in the hard dense loop
approximation, the potential $V$ in the model is given in 
Eq.~(\ref{V_exp}). However, it is very hard to perform
explicitly the calculations of the term with the fermion trace in  
$V$ in the case of nonzero gluon condensates (\ref{VEV-g}).  

The GL approach can help to overcome this difficulty
in studies of nearcritical dynamics
in systems with a second order phase transition. In this
approach, we need to pick up those local
operators ${\cal O}_n$
with the mass dimension four and less which are 
invariant under the symmetry of a system. In our case, it is  
\begin{equation}
 SU(2)_c \times
 \tilde{U}(1)_{\rm em} \times
 [U(1)_{\tau^3_L} \times U(1)_{\tau^3_R}]_{\chi} \times SO(3)_{\rm rot}
 \label{sym-2SC1}
\end{equation}
(see Eq. (\ref{sym-2SC})).
The GL effective action and Lagrangian, 
$S_{\rm eff}$ and ${\cal L}_{\rm eff}$,
are written in terms of the sum of these operators:
\begin{equation}
  S_{\rm eff}[A_\mu^a;\mu,\mu_e,\Delta] = \int dx^4
  {\cal L}_{\rm eff}[A_\mu^a;\mu,\mu_e,\Delta], \quad 
 {\cal L}_{\rm eff} = \sum_n K_n {\cal O}_n.
\end{equation}
The part of the Lagrangian without derivative terms 
yields the effective potential $V_{\rm eff}$.
By analyzing the structure of the
${\cal L}_{\rm eff}$, we can determine
the relevant terms before providing the 
calculations of the full potential $V$ (\ref{V_exp}).
It essentially reduces the amount of the work.

Let us consider the GL effective Lagrangian
in the gauged NJL
model. Because the $SU(2)_c$ in Eq. (\ref{sym-2SC1})
is a gauge symmetry,
the building blocks of ${\cal L}_{\rm eff}$ are
\begin{equation}
  \phi_0, \quad \phi_j, \quad {\cal D}_0, \quad {\cal D}_j,  \quad
  f_{0j}, \quad f_{jk}, \quad A_j^{(8)},
\label{blocks}
\end{equation}
where the indices $j$ and $k$ represent spatial components.
The coefficients $K_n$
of the operators are functions of $\Delta$, $\mu$,
$\mu_e$, and $\mu_8$. But when we expand the effective action with
respect to $\mu_8$, one should add the field $A_0^{(8)}$:
\begin{equation}
  \phi_0, \quad \phi_j, \quad {\cal D}_0, \quad {\cal D}_j,  \quad
  f_{0j}, \quad f_{jk}, \quad A_0^{(8)}, \quad A_j^{(8)},
\label{blocks1}
\end{equation}
and now the the coefficients of the operators are functions 
of $\Delta$, $\mu$, and $\mu_e$. For completeness, we will
describe the larger set of the operators expressed through
building blocks (\ref{blocks1}).

Up to the mass-dimension four, we find the operators without
the singlet field $A_\mu^{(8)}$,
\begin{subequations}
\label{op}
\begin{equation}
  \phi^\dagger_0 \phi_0, \quad \phi^\dagger_j \phi^j,
  \label{block2}
\end{equation}
\begin{equation}
  \phi^\dagger_0 (i{\cal D}_0 \phi_0), \quad
  \phi^\dagger_j (i{\cal D}_0 \phi^j), \quad
  \phi_0^\dagger (i{\cal D}_j \phi^j),
  \label{block3}
\end{equation}
\begin{equation}
  (\phi^\dagger_0 \phi_0)^2, \quad
  (\phi^\dagger_0 \phi_0)(\phi^\dagger_j \phi^j), \quad
  (\phi^\dagger_0 \phi^j)(\phi^\dagger_j \phi_0), \quad
  (\phi^\dagger_j \phi^j)^2, \quad
  (\phi^\dagger_j \phi^k)(\phi^\dagger_k \phi^j), \quad
  (\phi^\dagger_0 \phi^j)(\phi^\dagger_0 \phi_j), \quad
  (\phi^{j\dagger} \phi_k)(\phi^\dagger_j \phi^k),
  \label{block4-1}
\end{equation}
\begin{equation}
 |{\cal D}_0 \phi_0|^2, \quad |{\cal D}_j \phi_0|^2, \quad
 |{\cal D}_0 \phi_j|^2, \quad  |{\cal D}_j \phi_k|^2, \quad
  \phi_j^\dagger \left(\, ( {\cal D}_i{\cal D}^i\delta^{jk}
                + \frac{1}{2}\{{\cal D}^j,{\cal D}^k\})\phi_k\right), \quad
  \phi_0^\dagger (\{{\cal D}^j,{\cal D}^0\} \phi_j),
  \label{block4-2}
\end{equation}
and
\begin{equation}
  \tr(f_{0j}f^{0j}), \quad   \tr(f_{jk}f^{jk}), \quad
  i\phi_j^\dagger f^{jk}\phi_k, \quad
  \phi_0^\dagger f^{0j} \phi_j.
  \label{block4-3}
\end{equation}
\end{subequations}
Including $A_\mu^{(8)}$,
we also get the following operators,
\begin{subequations}
\label{op2}
\begin{equation}
  \phi^\dagger_0 \phi_0 A_0^{(8)}, \quad \phi^\dagger_j 
\phi^j A_0^{(8)}, \quad
  \phi^\dagger_0 \phi_0 (A_0^{(8)})^2, 
\quad \phi^\dagger_j \phi^j (A_0^{(8)})^2,\quad
  \phi^\dagger_0 \phi_0 \partial_0 A_0^{(8)}, \quad
  \phi^\dagger_j \phi^j \partial_0 A_0^{(8)}, \quad
  \phi^\dagger_0 \phi^j \partial_j A_0^{(8)},
  \label{block2-mu8}
\end{equation}
\begin{equation}
  \phi^\dagger_0 (i{\cal D}_0 \phi_0) A_0^{(8)}, \quad
  \phi^\dagger_j (i{\cal D}_0 \phi^j) A_0^{(8)}, \quad
  \phi_0^\dagger (i{\cal D}_j \phi^j) A_0^{(8)},
  \label{block3-mu8}
\end{equation}
\end{subequations}
and
\begin{subequations}
\label{op3}
\begin{equation}
  \phi^\dagger_0 \phi^j A_j^{(8)} , \quad
  \phi^\dagger_0 \phi^j A_j^{(8)} A_0^{(8)} , \quad
  \phi_0^\dagger \phi_0 (A_j^{(8)})^2,  \quad
  \phi_j^\dagger \phi^j (A_k^{(8)})^2,  \quad
  \phi^{j\,\dagger} \phi^k A_j^{(8)} A_k^{(8)}, \quad
  \phi^\dagger_0 \phi^j \partial_0 A_j^{(8)} , \quad
  \phi_j^\dagger \phi^k \partial^j A_k^{(8)},  \quad
  \label{block2-a}
\end{equation}
\begin{equation}
  \phi_0^\dagger \phi_0 \partial^j A_j^{(8)},  \quad
  \phi_j^\dagger \phi^j \partial^k A_k^{(8)},  \quad
  A_j^{(8)} \phi^\dagger_0 (i{\cal D}_0 \phi^j), \quad
  A_j^{(8)} \phi^\dagger_0 (i{\cal D}^j \phi_0), \quad
  A_k^{(8)} \phi^\dagger_j (i{\cal D}^k \phi^j), \quad
  A_k^{(8)} \phi^\dagger_j (i{\cal D}^j \phi^k). \quad
  \label{block3-a}
\end{equation}
\end{subequations}
The operators including only the singlet field $A_\mu^{(8)}$ are
\begin{equation}
  A_0^{(8)}, \quad (A_0^{(8)})^2, \quad   (A_j^{(8)})^2, \quad
  A_0^{(8)}(A_j^{(8)})^2, \quad
  \mbox{etc.} \, .
  \label{op4}
\end{equation}

The effective Lagrangian is given by the sum of the above operators.
It is understood that the 
hermitian conjugate operators are also included in ${\cal L}_{\rm eff}$,
and some coefficients can
be complex. The following remarks are in order. a) As has to be,
the sets (\ref{op}), (\ref{op2}), and (\ref{op3}) do not
contain operators corresponding to
the Debye and Meissner screening mass terms for 
the $SU(2)_c$ gauge gluons.
b)We do not consider the parity violating term such as
$\epsilon^{ijk}\phi_i^\dagger f_{0j} \phi_k \, ,\mbox{etc.}$.
c) It is easy to check
in the hard dense loop approximation
(with the one loop fermion contributions) that
due to the structure of the fermion-antifermion-gluon vertex,
a $n$-point vertex of gluons has an even number of 
the $\epsilon^b$-type vertices. Therefore we do not consider terms as
$\epsilon^{\alpha\beta}(\phi_j)_\alpha (i{\cal D}_0 \phi^j)_{\beta}
\, ,\mbox{etc.}$,
where $\alpha,\beta$ denote the $SU(2)_c$ indices.

We are ready to show that the GL effective potential has
the form presented in Eq. (\ref{LG-pot}). Indeed, now we know
all relevant and marginal operators which can be constructed
for the fields $B, C$ and $D$ defined in Eq. (\ref{BCD}).
They are: $\phi_j^\dagger \phi^j$ corresponding to $B^2$,
$\phi_j^\dagger (i{\cal D}_0\phi^j)$ to
$DB^2$, $|{\cal D}_j\phi_k|^2$ to $B^2C^2$,
$|{\cal D}_0 \phi_j|^2$ to $B^2D^2$, $\tr(f_{0j}f^{0j})$ to
$C^2D^2$, and $(\phi_j^\dagger \phi^j)^2$, $|\phi_j^\dagger \phi^k|^2$ 
and $(\phi_j^\dagger \phi^k)^2$ corresponding to $B^4$. 
Since there are no other relevant and marginal operators
which can be constructed from these three fields, we are led to
the form in Eq. (\ref{LG-pot}).

To obtain all of the coefficients $K_n$
of the operators is a very hard task.
Only some of them are known.
The coefficients of $\phi_0^\dagger \phi_0$ , $(A_0^{(8)})^2$,
and $\phi_j^\dagger \phi^j$, $(A_j^{(8)})^2$ 
correspond to the Debye and Meissner screening masses,
respectively~\cite{Huang:2004bg}.
The coefficients of the operators 
$\phi_j^\dagger (i{\cal D}_0\phi^j)$, $|{\cal D}_j\phi_k|^2$,
and $\tr(f_{0j}f^{0j})$ correspond to the parameters 
$T_{DB}$, $\lambda_{BC}$, and
$\lambda_{CD}$, respectively, in the GL potential in the gluonic phase
used in Sec. \ref{gluon} (see
Eq. (\ref{V_min})). These parameters are calculated in Appendix \ref{B}.
Without $\delta\mu$, the three and four point gluon vertices have been
obtained in Ref.~\cite{Casalbuoni:2001ha}.
The important consequences of those calculations
are that the coefficients of
the three point vertex operators 
$\phi_0^\dagger (i{\cal D}_0\phi^0)$,
$\phi_j^\dagger (i{\cal D}_0\phi^j)$, and 
$\phi_0^\dagger (i{\cal D}_j\phi^j)$
are vanishing at the order of $\mu^2$, (i.e., they 
$\sim {\cal O}(\mu)$),
and that four point vertex operators such as
$|{\cal D}_j\phi_k|^2$, $\tr(f_{0j}f^{0j})$, and
$A_k^{(8)} \phi_j^\dagger(i{\cal D}^k \phi^j)$ are of order $\mu^2$.
For their concrete expressions, see Ref.~\cite{Casalbuoni:2001ha}.
The results of our calculations in Appendix \ref{B} are
of course consistent with them.

The GL approach can be applied only to those symmetry breaking
samples, discussed
in Subsec. \ref{sec9}, which are connected with a second order phase
transition. Therefore, it cannot be applied both to the sample
A (the single plane-wave LOFF state) and the sample B
(the gluonic cylindrical phase II), which are related to
a first order phase transition (see Refs. 
\cite{Alford:2000ze,Giannakis:2004pf,Gorbar:2005tx} and
the discussion in Subsec. \ref{sec9}). 
\footnote{Note that as
follows from Refs. \cite{Huang:2004bg,Gorbar:2006up},
the instabilities at $\bar{\Delta} = \delta\mu$ in the 
g2SC phase (not considered in this paper), also 
cannot be connected with a conventional second order phase
transition.} 
On the other hand, as was shown in Sec. \ref{gluon}, 
the GL approach consistently describes the gluonic
cylindrical phase I, assigned to the symmetry breaking
sample C 
(a detailed analysis leading to this conclusion is considered in this
subsection below). The discussion  of the dynamics in
the symmetry breaking sample D 
(the gluonic CSL phase) is beyond the scope of this paper. 
 
To demonstrate the power of the GL approach, we will apply
it to prove a self-consistency of the ansatz for the gluonic
phase we used in Sec. \ref{gluon}. 
In Subsec. \ref{sec7}, a tadpole mechanism for producing
VEVs of gluon fields from triple vertices was discussed.
Let us describe it in more detail.
One can always divide a field into the VEV part
and the fluctuation one,
\begin{subequations}
\label{fluctuation}
\begin{eqnarray}
A_\mu^{(\ell)} &=& \VEV{A_\mu^{(\ell)}} + a_\mu^{(\ell)}, 
\quad (\ell = 1,2,3), \\
\phi_\mu &=& \VEV{\phi_\mu} + \zeta_\mu, \\
A_\mu^{(8)}  &=& \VEV{A_\mu^{(8)}} + a_\mu^{(8)} \, .
\end{eqnarray}
\end{subequations}
Substituting expressions ~(\ref{fluctuation}) 
for {\it all} fields
into the GL potential
expressed through operator 
sets (\ref{op})--(\ref{op4}), 
we obtain the VEV part, the tadpole part, which is
linear with respect to 
fluctuations, plus higher order terms.
The gap equation is the stationary condition for 
the effective potential and it is equivalent to 
vanishing 
the tadpole part. We emphasize that this condition is stronger than
solving the gap equation for a given ansatz by taking
into the consideration only the fluctuations 
of the fields in the ansatz and ignoring the fluctuations 
of other ones.

The condition of vanishing the tadpole contribution has to be correct 
for any consistent ansatz used for VEVs: its violation indicates
that the ansatz is inconsistent.
While for the {\it maximal} ansatz, corresponding to a chosen
symmetry breaking pattern, 
the violation would imply that the pattern cannot be realized in
the model at all, for  
a non-maximal one, this could just mean that the latter  
is not closed and one should add  
new VEVs for restoring its consistency. 

As a concrete example, let us consider the ansatz with 
$\VEV{A_0^{(8)}} \ne 0$,
$\VEV{A_x^{(4)}} \ne 0$, and $\VEV{A_y^{(8)}} \ne 0$,
which was discussed in Ref.~\cite{Fukushima:2006su},
and show that it
is {\it not} self-consistent. The point is that
the operators
$A_k^{(8)} \phi_j^\dagger(i{\cal D}^k \phi^j)$ and 
$\phi_j^\dagger (i{\cal D}_0 \phi^j)$ yield 
the tadpole contributions $a_y^{(3)} \VEV{A_y^{(8)}}\VEV{A_x^{(4)}}^2$ 
and $a_0^{(3)}\VEV{A_x^{(4)}}^2$, respectively. 
Because there are no other tadpole terms for $a_y^{(3)}$ and $a_0^{(3)}$,
the condition of
vanishing the tadpole contribution leads to $\VEV{A_x^{(4)}} = 0$,
which is inconsistent with the original ansatz.
If we started from the ansatz with $\VEV{A_0^{(3)}} \ne 0$, 
$\VEV{A_0^{(8)}} \ne 0$, $\VEV{A_y^{(3)}} \ne 0$, 
$\VEV{A_x^{(4)}} \ne 0$, and $\VEV{A_y^{(8)}} \ne 0$, new
tadpole contributions for $a_y^{(3)}$ and $a_0^{(3)}$ would appear.
They will either allow to cancel the previous tadpole contributions 
without generating new ones or  
this modified ansatz is also inconsistent and should be in its turn
extended.
This process will eventually lead either to a consistent ansatz or
to the conclusion 
that this symmetry breaking pattern is not realized in the model.
\footnote{This example corresponds to one of the non-maximal
ans\"atze 
for the symmetry breaking sample E considered in the previous subsection. 
For the choice of the VEVs in the phase E used in the present paper,
$\VEV{A_y^{(3)}} \ne 0$, $\VEV{A_x^{(4)}} \ne 0$ and 
$\VEV{A_y^{(8)}} \ne 0$ in the ansatz should be replaced by 
$\VEV{A_z^{(3)}} \ne 0 $, $\VEV{A_y^{(6)}} \ne 0$ and 
$\VEV{A_z^{(8)}} \ne 0$.}

Without calculating the coefficients of the operators 
and solving the gap equations, this check helps to
pick up a self-consistent ansatz.
By estimating the size of the coefficients, one can also specify 
suppressed VEVs, as we will see below. The latter can be
useful for simplifying the ansatz we work with.

Let us now turn to the gluonic phase in Sec. \ref{gluon}. As was
shown in Subsec.VI A, the phase C, 
having the same symmetry
structure as the gluonic cylindrical phase I, can contain other 
VEVs than $B=g\VEV{A_z^{(6)}}$, $C=g\VEV{A_z^{(1)}}$,  
$D=g\VEV{A_0^{(3)}}$, and $\mu_8=\sqrt{3}/2 g\VEV{A_0^{(8)}}$ 
used in Sec.V.
Is the ansatz including only $B$, $C$, $D$, and 
$\mu_8$ self-consistent?
Based on the GL approach, we will show that the answer to
this question is affirmative. Actually, we will show that,
strictly speaking, the self-consistent minimal ansatz for
the gluonic cylindrical phase I is
\begin{equation}
\VEV{A_z^{(6)}}, \VEV{A_z^{(1)}}, \VEV{A_0^{(3)}}, \VEV{A_0^{(8)}}, \quad
\mbox{and} \quad \VEV{A_0^{(4)}}, \VEV{A_0^{(5)}} \, .
\label{min}
\end{equation} 
However, it will be shown that the additional VEVs 
$\VEV{A_0^{(4)}}$ and $\VEV{A_0^{(5)}}$ are
suppressed in the vicinity of the critical point.

For $B,C,D \ne 0$, the tadpole contributions 
for $a_0^{(4,5)}$ 
come from the operators 
$\phi_0^\dagger (i{\cal D}_j \phi^j)$ and 
$\phi_0^\dagger f^{0j} \phi_j$ in the GL effective Lagrangian
with, say, coefficients $K_1$ and $K_2$, respectively. The
condition of
vanishing the tadpole part yields $D \sim -K_1/K_2$.
On the other hand, the solution of the gap equations is given in
Eq.~(\ref{appr}). Since $M_B^2$ and $T_{DB}$ are the coefficients
of different operators, $\phi_j^\dagger \phi^j$ and 
$\phi_j^\dagger (i{\cal D}_0 \phi^j)$, respectively, it is hard
to expect a ``magic'' 
relation (fine tuning) 
among $K_1$, $K_2$, $M_B^2$ and $T_{DB}$ coefficients, which would
lead to the same solution for $D$ from the gap equation.
Therefore, one should introduce VEVs for $A_0^{(4,5)}$ in this case.

Noting that $A_0^{(4,5)}$ have nonzero Debye mass 
$m_{D,4} \sim {\cal O}(\mu)$,
one can then estimate the solution of the gap equations for 
$\VEV{A_0^{(4,5)}}$ as
$\VEV{A_0^{(4)}}, \VEV{A_0^{(5)}} \sim K_1 BC/m_{D,4}^2 + K_2BCD/m_{D,4}^2$.
It is obvious that near the critical point, where
$B, C, D \to 0$, the contribution of the first operator dominates.
Without $\delta\mu$, the coefficients $K_1$ and $K_2$ were
calculated in hard dense loop approximation in
Ref.~\cite{Casalbuoni:2001ha}. It was shown there that while
$K_2$ is of order $\mu^2/\Delta^2$, $K_1$ is vanishing
in this leading approximation.
It implies that  $K_1 \sim {\cal O}(\mu)$ or less,
and thereby we conclude that
the contribution of this operator to the effective potential is the
term $B^2 C^2$ with the coefficient $\sim {\cal O}(1)$ at most.
On the other hand, the coefficient of the $B^2 C^2$ term discussed
in Sec.V is $\lambda_{BC}$ in Eq.~(\ref{bc}):
$\lambda_{BC} \sim {\cal O}(\mu^2/\Delta^2)$.
Therefore, the contributions of $\VEV{A_0^{(4)}}$ and
$\VEV{A_0^{(5)}}$ can be ignored in the vicinity of the critical
point. In addition to this, one can show that with the ansatz (\ref{min}) 
for VEVs, there are no 
potentially dangerous operators in Eqs. (\ref{op})--(\ref{op4}) which would 
lead to tadpole contributions for $A_z^{(8)}$ and other additional fields
from the maximal ansatz absent in Eq. (\ref{min}). This allows us
to conclude that the ansatz used in Sec. \ref{gluon} is
self-consistent indeed. 
The above example clearly shows the power of the GL approach
and its relevance for the dynamics with vector condensates of
gluons.
\footnote{Another example of this analysis is for the gluonic
CSL phase D. One can show that with the GL operators 
in Eqs. (118)-(121), there are two self consistent 
ans\"atze in that case:
the maximal one and the maximal one minus the VEV of the
field $A_{0}^{(2)}$.}

\section{Conclusion}
\label{conclusion}

The gluonic phases yield an example of dynamics in gauge models
with matter in which the Higgs mechanism is
provided by condensates of gauge (or gauge plus scalar) fields.
Because VEVs of spatial components of vector fields break
the rotational symmetry, it is natural to have a spontaneous
breakdown both of external and internal symmetries in this
case. Dynamics in such systems are quite
sophisticated. 
The existence of exotic hadrons in the gluonic phases is
especially intriguing.

What could be directions for future studies
in these phases? It is evident that it would be interesting 
to consider the spectrum of light collective excitations there.
The results in Refs. \cite{sigmamodel,Gorbar:2005pi}, obtained
in the gauged $\sigma$-model with a chemical potential for
hypercharge (briefly
discussed in Sec. \ref{sec2}), suggest that the spectrum should be
rich, containing, in particular, 
gapless NG modes, rotonlike and vortexlike excitations.

Another interesting direction would be to clarify whether the symmetry
breaking patterns considered in Subsec. \ref{sec9}, can be 
realized as stable, metastable or unstable ground states in
dense QCD. Recent results \cite{Buchel:2006aa}, showing that the
landscape of such ground states in the gauged $\sigma$-model with 
a chemical potential for hypercharge is rich, are
encouraging. 

It is clear that it would be worth to figure out whether
phases with vector condensates of gluons could exist in dense
matter with three quark flavors. Recently, this possibility has
been mentioned in 
Refs. \cite{Gorbar:2006up,Fukushima:2006su,Zhang:2006rp,Gerhold:2006np}.

It has been recently revealed that a strong enough magnetic field
can influence the phase structure in dense quark matter
\cite{Ferrer:2005vd}. 
\footnote{This phenomenon has some similarities
with the phenomenon of the magnetic catalysis in vacuum field
theories \cite{Gusynin1995PRD}, in particular, in the vacuum QCD 
\cite{Miransky:2002rp}.}
It would be interesting to study this phenomenon in gluonic
phases, especially because one can expect the existence of vortices
there \cite{Gorbar:2005pi,Ferrer:2006ie}.

Last but not least, it would be worth searching for a realization
of a gluoniclike phase in condensed matter.
Recently,
there has been a considerable interest in systems with coexisting
order parameters (such as
high $T_c$ superconductors) in condensed matter \cite{Sachd}. 
Generating vector condensates is a natural way of creating
such systems (for example, in the gluonic cylindrical phase I, 
electric superconductivity
coexists with spontaneous rotational symmetry breaking).   

\acknowledgments

We acknowledge useful discussions with Alex Buchel, Junji Jia, and
Igor Shovkovy. The
work of V.A.M. was supported by the Natural Sciences and Engineering
Research Council of Canada. 
He is grateful to Prof. Taichiro Kugo and Prof. Teiji Kunihiro 
for their warm hospitality during his stay
at Yukawa Institute for Theoretical Physics, Kyoto University. 
Discussions during the YKIS2006 ``New Frontiers in QCD'' were
useful for completing this work.

\appendix

\section{Structure of nearcritical solution in gluonic phase}
\label{A}

In this Appendix, we analyze the structure of the nearcritical
solution with $B \ne 0$, $C \ne 0$, and $D \ne 0$ discussed
in Subsec. \ref{sec5}. Eqs.~(\ref{gap-eq-B})-(\ref{gap-eq-D}) 
yield
\begin{equation}
  D = - \frac{M_B^2}{3T_{DB}}
  + \left(\,\frac{\lambda_{B}\lambda_{CD}}{\lambda_{BC}}-2\lambda_{BD}
         \,\right) \frac{D^2}{3T_{DB}} , \quad
  B^2 = -\frac{\lambda_{CD}}{\lambda_{BC}}D^2 , \quad
  C^2 =  \frac{T_{DB}}{\lambda_{BC}}D
        +\frac{\lambda_{BD}}{\lambda_{BC}}D^2 .
  \label{sol_BCD}
\end{equation}
On the other hand, Eq.~(\ref{parameters}) yields:
\begin{subequations}
\label{gap_Delta2}
\begin{eqnarray}
  \frac{\partial V_\Delta}{\partial \Delta} &=& -\frac{1}{2}B^2
  \frac{\partial M_B^2}{\partial \Delta}
   - D B^2 \frac{\partial T_{DB}}{\partial \Delta}
   -\frac{1}{2}B^2 C^2 \frac{\partial \lambda_{BC}}{\partial \Delta}
   -\frac{1}{2}C^2 D^2 \frac{\partial \lambda_{CD}}{\partial \Delta}
   -\frac{1}{2}B^2 D^2 \frac{\partial \lambda_{BD}}{\partial \Delta}
   -\frac{1}{4}B^4 \frac{\partial \lambda_B}{\partial \Delta}, \\[3mm]
  \frac{\partial V_\Delta}{\partial \mu_e} &=& -\frac{1}{2}B^2
  \frac{\partial M_B^2}{\partial \mu_e}
   - D B^2 \frac{\partial T_{DB}}{\partial \mu_e}
   -\frac{1}{2}B^2 C^2 \frac{\partial \lambda_{BC}}{\partial \mu_e}
   -\frac{1}{2}C^2 D^2 \frac{\partial \lambda_{CD}}{\partial \mu_e}
   -\frac{1}{2}B^2 D^2 \frac{\partial \lambda_{BD}}{\partial \mu_e}
   -\frac{1}{4}B^4 \frac{\partial \lambda_B}{\partial \mu_e}, \\[3mm]
  \frac{\partial V_\Delta}{\partial \mu_8} &=& -\frac{1}{2}B^2
  \frac{\partial M_B^2}{\partial \mu_8}
   - D B^2 \frac{\partial T_{DB}}{\partial \mu_8}
   -\frac{1}{2}B^2 C^2 \frac{\partial \lambda_{BC}}{\partial \mu_8}
   -\frac{1}{2}C^2 D^2 \frac{\partial \lambda_{CD}}{\partial \mu_8}
   -\frac{1}{2}B^2 D^2 \frac{\partial \lambda_{BD}}{\partial \mu_8}
   -\frac{1}{4}B^4 \frac{\partial \lambda_B}{\partial \mu_8} .
\end{eqnarray}
\end{subequations}
Because the coefficients $M_B^2$, $T_{DB}$, etc. in Eq.~(\ref{sol_BCD})
are the functions of $\Delta$, $\mu_e$ and $\mu_8$,
Eqs.~(\ref{sol_BCD}) and (\ref{parameters}) constitute a coupled system
of six equations. However, as will be shown below, near the critical point,
the equations (\ref{sol_BCD}) and 
(\ref{parameters}) decouple.

Near the critical point, $M_B^2 \approx 0$,
one can expand $\Delta$, $\mu_e$ and $\mu_8$ around
$B=C=D=0$:
\begin{eqnarray}
  \Delta &=& \bar{\Delta} + \xi_\Delta, \\
  \mu_e &=& \bar{\mu}_e + \xi_e, \\
  \mu_8 &=& \bar{\mu}_8 + \xi_8 ,
\end{eqnarray}
where the bar-quantities correspond to
the 2SC solution with $B=C=D=0$.
By definition of the bar-quantities,
the following stationary conditions are satisfied:
\begin{equation}
  \frac{\partial V_\Delta}{\partial \Delta}
   \bigg|_{\Delta=\bar{\Delta},\mu_e=\bar{\mu}_e,\mu_8=\bar{\mu}_8} = 0, \quad
  \frac{\partial V_\Delta}{\partial \mu_e}
   \bigg|_{\Delta=\bar{\Delta},\mu_e=\bar{\mu}_e,\mu_8=\bar{\mu}_8} =  0, \quad
  \frac{\partial V_\Delta}{\partial \mu_8}
  \bigg|_{\Delta=\bar{\Delta},\mu_e=\bar{\mu}_e,\mu_8=\bar{\mu}_8} = 0.
\end{equation}
They yield
\begin{eqnarray}
  \bar{\mu}_e &=& \frac{3}{5}\mu-\frac{2}{5}\bar{\mu}_8, \label{2SC-1} \\[3mm]
  \left(\frac{3}{5}\mu-\frac{2}{5}\bar{\mu}_8\right)^3-\frac{2}{9}
  \left(\frac{9}{10}\mu+\frac{2}{5}\bar{\mu}_8\right)^3 &=&
  \frac{1}{3}\bar{\Delta}^2\left(\frac{9}{10}\mu+\frac{2}{5}\bar{\mu}_8\right)
  \left(\,\ln\frac{4\Lambda^2}{\bar{\Delta}^2}-2\,\right),
  \label{2SC-2} \\[3mm]
  \left[\,\left(\frac{9}{10}\mu+\frac{2}{5}\bar{\mu}_8\right)^2
          -\frac{1}{2}\bar{\Delta}^2\,\right]
  \ln\frac{4\Lambda^2}{\bar{\Delta}^2} &=&
  \frac{\pi^2}{4G_\Delta} - \Lambda^2
   + 3\left(\frac{9}{10}\mu+\frac{2}{5}\bar{\mu}_8\right)^2
   - \frac{1}{2}\bar{\Delta}^2 . \label{2SC-3}
\end{eqnarray}
Note that the bar-quantities $\bar{\Delta}$, $\bar{\mu}_e$
and $\bar{\mu}_8$ are uniquely determined
by the theoretical parameters $G_\Delta,\Lambda,\mu$
in the gauged NJL model.
Expanding the left hand side in
Eq.~(\ref{gap_Delta2}) around the bar-quantities,
we obtain the gap equation for $\xi_\Delta$, $\xi_e$ and $\xi_8$
expressed in a matrix form as
\begin{equation}
  \left(
  \begin{array}{ccc}
  {\displaystyle\frac{\partial^2 V_\Delta}{\partial \Delta^2}} &
  {\displaystyle\frac{\partial^2 V_\Delta}{\partial \mu_e \partial \Delta}} &
  {\displaystyle\frac{\partial^2 V_\Delta}{\partial\mu_8\partial\Delta}}\\[3mm]
  {\displaystyle\frac{\partial^2 V_\Delta}{\partial \Delta \partial \mu_e}} &
  {\displaystyle \frac{\partial^2 V_\Delta}{\partial \mu_e^2}} &
  {\displaystyle\frac{\partial^2 V_\Delta}{\partial \mu_8\partial \mu_e}}\\[3mm]
  {\displaystyle\frac{\partial^2 V_\Delta}{\partial \Delta \partial \mu_8}} &
  {\displaystyle\frac{\partial^2 V_\Delta}{\partial \mu_e \partial \mu_8}} &
  {\displaystyle\frac{\partial^2 V_\Delta}{\partial \mu_8^2}}
  \end{array}
  \right)
  \left(
  \begin{array}{@{}c@{}}
  \xi_\Delta \\[8mm] \xi_e \\[8mm] \xi_8
  \end{array}
  \right) = -\frac{1}{2}B^2
  \left(
  \begin{array}{c}
  {\displaystyle\frac{\partial M_B^2}{\partial \Delta}} \\[5mm]
  {\displaystyle\frac{\partial M_B^2}{\partial \mu_e}} \\[5mm]
  {\displaystyle\frac{\partial M_B^2}{\partial \mu_8}}
  \end{array}
  \right) + {\cal O}(DB^2) + {\cal O}(B^2 C^2) + {\cal O}(C^2 D^2),
  \label{gap-matrix}
\end{equation}
where all derivatives are calculated at
$\Delta=\bar{\Delta}, \mu_e=\bar{\mu}_e$, and $\mu_8=\bar{\mu}_8$.

Combining Eq.~(\ref{sol_BCD}) with Eq.~(\ref{gap-matrix}),
we find the following approximate solution:
\begin{equation}
  B_{\rm sol} \simeq \frac{-\bar{M}_B^2}{3|\bar{T}_{DB}|}
           \sqrt{\frac{-\bar{\lambda}_{CD}}{\bar{\lambda}_{BC}}}, \quad
  C_{\rm sol} \simeq  \sqrt{\frac{-\bar{M}_B^2}{3\bar{\lambda}_{BC}}}, \quad
  D_{\rm sol} \simeq \frac{-\bar{M}_B^2}{3\bar{T}_{DB}}, \quad
  \xi_\Delta^{\rm sol}, \xi_e^{\rm sol}, \xi_8^{\rm sol} \sim
  {\cal O}((\bar{M}_B^2)^2),
  \label{app}
\end{equation}
where higher order terms in $\bar{M}_B^2$ were neglected.
Here all the coefficients $\bar{M}_B^2$, $\bar{T}_{DB}$,
$\bar{\lambda}_{BC}$, and $\bar{\lambda}_{CD}$ are expressed
through
the 2SC values $\bar{\Delta}$, $\bar{\mu}_e$ and $\bar{\mu}_8$.
This solution exists when
\begin{equation}
  \bar{M}_B^2 < 0 , \quad \bar{\lambda}_{BC} > 0, \quad
  \bar{\lambda}_{CD} < 0.
\end{equation}
Note that in Eq.~(\ref{app}) the convention $B_{\rm sol} > 0$ and
$C_{\rm sol} > 0$ is chosen.

Substituting solution (\ref{app}) 
into GL potential (\ref{LG-pot}),
we find that its gluonic part $V_{\rm eff}-V_\Delta$ 
consists of leading terms $\sim {\cal O}((M_B^2)^3)$
and subleading terms $\sim {\cal O}((M_B^2)^4)$.
In particular,
the deviation of $V_\Delta$ from that in the 2SC solution is
of subleading order $\sim {\cal O}((M_B^2)^4)$. Indeed,
the second order of the Taylor expansion for $V_\Delta$ yields
\begin{eqnarray}
\lefteqn{
   V_\Delta(\Delta^{\rm sol},\mu_e^{\rm sol},\mu_8^{\rm sol})
 - V_\Delta(\bar{\Delta},\bar{\mu}_e,\bar{\mu}_8)
} \nonumber \\[5mm] &&
 =\frac{1}{2}
  \left(
  \begin{array}{ccc}
  \xi_\Delta^{\rm sol} & \xi_e^{\rm sol} & \xi_8^{\rm sol}
  \end{array}
  \right)
  \left(
  \begin{array}{ccc}
  {\displaystyle\frac{\partial^2 V_\Delta}{\partial \Delta^2}} &
  {\displaystyle\frac{\partial^2 V_\Delta}{\partial \mu_e \partial \Delta}} &
  {\displaystyle\frac{\partial^2 V_\Delta}{\partial \mu_8\partial\Delta}}\\[3mm]
  {\displaystyle\frac{\partial^2 V_\Delta}{\partial \Delta \partial \mu_e}} &
  {\displaystyle \frac{\partial^2 V_\Delta}{\partial \mu_e^2}} &
  {\displaystyle\frac{\partial^2 V_\Delta}{\partial \mu_8\partial\mu_e}}\\[3mm]
  {\displaystyle\frac{\partial^2 V_\Delta}{\partial \Delta \partial \mu_8}} &
  {\displaystyle\frac{\partial^2 V_\Delta}{\partial \mu_e \partial \mu_8}} &
  {\displaystyle\frac{\partial^2 V_\Delta}{\partial \mu_8^2}}
  \end{array}
  \right)
  \left(
  \begin{array}{@{}c@{}}
  \xi_\Delta^{\rm sol} \\[8mm] \xi_e^{\rm sol} \\[8mm] \xi_8^{\rm sol}
  \end{array}
  \right) \\ &&
 = \frac{1}{8}B_{\rm sol}^4
  \left(
  \begin{array}{ccc}
  {\displaystyle\frac{\partial M_B^2}{\partial \Delta}} &
  {\displaystyle\frac{\partial M_B^2}{\partial \mu_e}}  &
  {\displaystyle\frac{\partial M_B^2}{\partial \mu_8}}
  \end{array}
  \right)
  \left(
  \begin{array}{ccc}
  {\displaystyle\frac{\partial^2 V_\Delta}{\partial \Delta^2}} &
  {\displaystyle\frac{\partial^2 V_\Delta}{\partial \mu_e \partial \Delta}} &
  {\displaystyle\frac{\partial^2 V_\Delta}{\partial \mu_8\partial\Delta}}\\[3mm]
  {\displaystyle\frac{\partial^2 V_\Delta}{\partial \Delta \partial \mu_e}} &
  {\displaystyle \frac{\partial^2 V_\Delta}{\partial \mu_e^2}} &
  {\displaystyle\frac{\partial^2 V_\Delta}{\partial \mu_8\partial\mu_e}}\\[3mm]
  {\displaystyle\frac{\partial^2 V_\Delta}{\partial \Delta \partial \mu_8}} &
  {\displaystyle\frac{\partial^2 V_\Delta}{\partial \mu_e \partial \mu_8}} &
  {\displaystyle\frac{\partial^2 V_\Delta}{\partial \mu_8^2}}
  \end{array}
  \right)^{-1}
  \left(
 \begin{array}{c}
  {\displaystyle\frac{\partial M_B^2}{\partial \Delta}} \\[5mm]
  {\displaystyle\frac{\partial M_B^2}{\partial \mu_e}} \\[5mm]
  {\displaystyle\frac{\partial M_B^2}{\partial \mu_8}}
  \end{array}
  \right),
\end{eqnarray}
where Eq. (\ref{gap-matrix}) was used (here all derivatives are calculated at
$\Delta=\bar{\Delta}, \mu_e=\bar{\mu}_e$, and $\mu_8=\bar{\mu}_8$).

Omitting the subleading terms, we arrive at the reduced
effective potential,
\begin{equation}
  \tilde{V}_{\rm eff} = V_\Delta (\bar{\Delta}, \bar{\mu}_e,\bar{\mu}_8)
  + \frac{1}{2}\bar{M}_B^2 B^2
  + \bar{T}_{DB} D B^2 + \frac{1}{2}\bar{\lambda}_{BC} B^2 C^2
  + \frac{1}{2}\bar{\lambda}_{CD} C^2 D^2 .
  \label{V_eff2}
\end{equation}
This potential is composed of two parts:
the ``constant''
2SC part $V_\Delta$, with frozen fermion parameters, and the dynamical
gluonic part.

\section{Coefficients in GL effective potential}
\label{B}

In this Appendix, we calculate the coefficients $\lambda_{BC}$,
$\lambda_{CD}$, and $T_{DB}$ of the marginal and relevant operators in
the reduced GL effective potential (\ref{V_min1}).
The coefficients $\lambda_{BC}$ and $\lambda_{CD}$ are connected
with the operators $|{\cal D}_j \phi_k|^2$ and $\tr(f_{0j}f^{0j})$,
respectively.
Therefore, they can be obtained from the kinetic terms
of the 4-7th and 1-3rd gluons.
Although the coefficient $T_{DB}$ of the triple vertex 
is connected with
the operator $\phi^\dagger_j (i{\cal D}_0 \phi^j)$,
it comes from a non-hard-dense-loop part, as we will see below.
Because of that, $T_{DB}$ will be directly calculated 
from the corresponding three-point vertex.
In passing,
the coefficient $\lambda_B$ in GL potential (\ref{LG-pot})
is connected with
the operator $(\phi_j^\dagger \phi^j)^2$
and cannot be reduced to a calculation of
a two-point function.
On the other hand, because the coefficient $\lambda_{BD}$ 
in (\ref{LG-pot}) is connected with the 
operator $|{\cal D}_0 \phi_j|^2$, it can be obtained by
taking  time-derivatives of the vacuum polarization tensor
for the 6th gluon.

One of the main difficulties in calculating these coefficients
is related to the presence of the gauge fields 
$C \equiv gA_z^{(1)}$ and $D  \equiv gA_0^{(3)}$ connected
with noncommutating color matrices $T^1$ and $T^3$.
In order to overcome it,
the following trick will be used.
We decompose the inverse 
Nambu-Gor'kov propagator (\ref{Sg-inv})
with gluon fields as 
\begin{equation}
  S_g^{-1}(X) = S_C^{-1}(X) + {\cal M}_{BD},
\end{equation}
where
\begin{equation}
  S_C^{-1}(X) = \left(
  \begin{array}{cc}
  i\fsl{\partial}+(\tilde{\mu}-\delta\mu\tau_3)\gamma^0 + CT^1\gamma^3
  & -i\varepsilon\epsilon^b\gamma_5\Delta \\
    -i\varepsilon\epsilon^b\gamma_5\Delta
  & i\fsl{\partial}-(\tilde{\mu}-\delta\mu\tau_3)\gamma^0 - CT^1\gamma^3
  \end{array}
  \right)
  \label{S-inv-C}
\end{equation}
and
\begin{equation}
 {\cal M}_{BD} \equiv
  \left(
  \begin{array}{cc}
  D T^3 \gamma^0 - B T^6 \gamma^3 & 0 \\
  0 & -D T^3 \gamma^0 + B T^6 \gamma^3
  \end{array}
  \right) ,
\end{equation}
with $X^\mu = (x^0,x^1,x^2,x^3)=(t,x,y,z)$ denoting space-time coordinates.
The point is that since
the diquark gap $\Delta$ is $SU(2)_c$-invariant, we
can remove the field $C = g A_z^{(1)}$ from 
$S_C(X)$
by using a $SU(2)_c$ gauge transformation.
[Note that since the field
strength is nonzero in the gluonic
phase,
it is impossible to remove all gauge fields
from the propagator $S_g(X)$]. In
fact, one can easily find this $SU(2)_c$ gauge transformation: 
\begin{equation}
  U(X) \equiv
  \left(
  \begin{array}{cc}
   e^{-i C T^1 x^3} & 0 \\
   0 & e^{i C T^1 x^3}
  \end{array}
  \right) 
  \label{def-U}
\end{equation}
transforms
the inverse $S_C^{-1}(X)$ in Eq.~(\ref{S-inv-C}) into
$S^{-1}(X)$ without $C$, i.e.,
\begin{equation}
  S_C^{-1}(X) = U^\dagger(X) S^{-1}(X) U(X).
  \label{S-inv2-C}
\end{equation}

Let us now expand the potential $V$ (\ref{V})
in the power series with respect to
$S_C(P)$ in momentum space:
\begin{equation}
  V  = V_\Delta + V_g
  +\sum_{n=1}^\infty \frac{(-1)^n}{2n}\int\frac{d^4 P}{i(2\pi)^4}
   \Tr (S_C {\cal M}_{BD})^n,
  \label{V_exp_gII}
\end{equation}
where the identity
\begin{equation}
  \Tr\ln S_C^{-1} = \Tr\ln (\,U S_C^{-1} U^\dagger\,) = \Tr\ln S^{-1} 
\end{equation}
was used.
Note that, by construction, the $V_g$ term is
$C$ independent.

In order to utilize expression (\ref{V_exp_gII}), we need to
calculate the propagator $S_C(P)$ in momentum space. 
From Eq. (\ref{S-inv2-C}) we get
\begin{equation}
  S_C(X) = U^\dagger(X) S(X) U(X).
  \label{gaugetransformation}
\end{equation}
Using then the relations
\begin{equation}
  e^{i C T^1 x^3} =
  \left(
  \begin{array}{cc}
   e^{iC' \sigma^1 x^3} & 0 \\
   0 & 1
  \end{array}
  \right)_c ,
  \quad
  e^{iC' \sigma^1 x^3}
  = \frac{1}{2}\left(\,e^{iC'x^3}+e^{-iC'x^3}\,\right)
   +\frac{1}{2}\sigma^1\left(\,e^{iC'x^3}-e^{-iC'x^3}\,\right),
\end{equation}
with
\begin{equation}
  C' \equiv \frac{1}{2}C,
\end{equation}
we find that the Fourier transform of $e^{iC' \sigma^1 x^3}$ 
is given by the sum of
two $\delta$-functions:
\begin{equation}
  \mbox{F.T.} e^{iC' \sigma^1 x^3} =
  \frac{1}{2}\left(\,\delta(p^3+C')+\delta(p^3-C')\,\right)
 +\frac{1}{2}\sigma^1\left(\,\delta(p^3+C')-\delta(p^3-C')\,\right).
 \label{ft-u}
\end{equation}
This implies that the Fourier transform of $U(X)$ (\ref{def-U}) 
is expressed through these $\delta$-functions. 
And since the Fourier transform of $S(X)$ is known, the Fourier
transform of $S_C(X)$ can be easily found from
Eqs. (\ref{gaugetransformation}) and (\ref{ft-u}):
\begin{equation}
  S_C(P) \equiv S_{rg}(P) \oplus S_b(P),
\end{equation}
with the red-green part
\begin{equation}
  S_{rg} (P) = \left(
  \begin{array}{cc}
  G^+_{rg} & \Xi^-_{rg} \\ \Xi^+_{rg} &  G^-_{rg}
  \end{array}
  \right) 
  \label{S_rg2}
\end{equation}
and the blue one
\begin{equation}
  S_{b}(P) = \left(
  \begin{array}{cc}
    (\fsl{P}+(\tilde{\mu}-\delta\mu\tau_3-\mu_8)\gamma^0)^{-1}
  & 0 \\ 0
  & (\fsl{P}-(\tilde{\mu}-\delta\mu\tau_3-\mu_8)\gamma^0)^{-1}
  \end{array}
  \right)
 =\left( \begin{array}{cc} G_b^+(P) & 0 \\ 0 & G_b^-(P) \end{array} \right).
  \label{S_b}
\end{equation}
While $G_{b}^{\pm}$ is given in Eq. (\ref{G_b}),
the matrix elements $G^{\pm}_{rg}$ and $\Xi^{\pm}_{rg}$ in
$S_{rg} (P)$ are connected with
the gapped part $S_\Delta(P)$,
\begin{equation}
  S_\Delta(P) = \left(
  \begin{array}{cc}
  G^+_\Delta & \Xi^-_\Delta \\ \Xi^+_\Delta &  G^-_\Delta
  \end{array}
  \right) 
  \label{S_del}
\end{equation}
with
\begin{equation}
  \Xi^\pm_\Delta (P) \equiv i\sigma^2
  \left(\begin{array}{cc}
    0 & \Xi_{12}^\pm \\ -\Xi_{21}^\pm & 0
  \end{array}\right)_f,
\end{equation}
where the quantities  $G^\pm_\Delta$, $G_b^\pm$, $\Xi_{12,21}$ are
given in Eqs.~(\ref{G_del}), (23), and (\ref{Xi}). 
In fact, $G^{\pm}_{rg}$ and $\Xi^{\pm}_{rg}$ are:
\begin{equation}
  G^\pm_{rg}(P) \equiv
  \frac{1}{2}\left(\,G_\Delta^\pm(P+P_{C})+G_\Delta^\pm(P-P_{C})\,\right)
 +\frac{1}{2}\left(\,\pm G_\Delta^\pm(P+P_{C})
                     \mp G_\Delta^\pm(P-P_{C})\,\right) \sigma^1,
\end{equation}
\begin{equation}
  \Xi^\pm_{rg}(P) \equiv
  \frac{1}{2}\left(\,\Xi_\Delta^\pm(P+P_{C})+
\Xi_\Delta^\pm(P-P_{C})\,\right)
 +\frac{1}{2}\left(\,\pm \Xi_\Delta^\pm(P+P_{C})
                     \mp \Xi_\Delta^\pm(P-P_{C})\,\right) \sigma^1,
\end{equation}
where  $P_C$ is the four-vector $(0, 0, 0, C/2)$.
Note that $\Xi_\Delta^\pm \sigma^1 = -\sigma^1 \Xi_\Delta^\pm$ and,
using Eq.~(\ref{G-Xi}), one can check that
$S^{-1}_{rg}(P)S_{rg}(P)=1$.

After taking the trace over the color indices,
we obtain the square term for $\lambda_{BC}$ and $\lambda_{CD}$,
\begin{eqnarray}
 \mathop{\Tr}_{c,f,s}(S_C {\cal M}_{BD})^2 &=& \phantom{+}
 \frac{1}{4}B^2 \mathop{\Tr}_{f,s} \bigg[\,
  \left(G_\Delta^+(P+P_{C})+G_\Delta^+(P-P_{C})\right)
\gamma^3 G_b^+(P) \gamma^3
 \nonumber \\ && \hspace*{2cm}\quad
 +\left(G_\Delta^-(P+P_{C})+G_\Delta^-(P-P_{C})\right)
\gamma^3 G_b^-(P) \gamma^3 \,\bigg]
 \nonumber \\ && +
 \frac{1}{2}D^2 \mathop{\Tr}_{f,s} \bigg[\,
  G_\Delta^+(P+P_{C}) \gamma^0 G_\Delta^+(P-P_{C}) \gamma^0
 \nonumber \\ && \hspace*{2cm} \quad
 +G_\Delta^-(P+P_{C}) \gamma^0 G_\Delta^-(P-P_{C}) \gamma^0 \,\bigg]
 \nonumber \\ && +
 \frac{1}{2}D^2 \mathop{\tr}_s \bigg[\,
  \Xi_{12}^-(P+P_{C}) \gamma^0 \Xi_{21}^+(P-P_{C}) \gamma^0
 +\Xi_{12}^-(P-P_{C}) \gamma^0 \Xi_{21}^+(P+P_{C}) \gamma^0
 \nonumber \\ && \hspace*{1.5cm}
 +\Xi_{21}^-(P+P_{C}) \gamma^0 \Xi_{12}^+(P-P_{C}) \gamma^0
 +\Xi_{21}^-(P-P_{C}) \gamma^0 \Xi_{12}^+(P+P_{C}) \gamma^0 \,\bigg] .
 \label{tr_2}
\end{eqnarray}
In fact, the $B^2$- and $D^2$-terms in Eq.~(\ref{tr_2})
correspond to the vacuum polarization tensors
$\tilde{\Pi}^{jj}_{44,55}$ and $\Pi_{11}^{00}$, respectively
(see Eqs.~(76) and (40) in the second paper in
Ref.~\cite{Huang:2004bg}.)
Therefore, the calculations of $\lambda_{BC}$ and $\lambda_{CD}$
are reduced to those of $\tilde{\Pi}^{jj}_{44,55}$ and $\Pi_{11}^{00}$.
We find
\begin{eqnarray}
\lambda_{BC} &=& \frac{1}{80\pi^2}\frac{\tilde{\mu}^2}{\Delta^2}
             \left[\,-1+8\frac{\delta\mu^2}{\Delta^2}
             \left(\,1-\frac{\delta\mu^2}{\Delta^2}\,\right)
            +\theta(\delta\mu-\Delta)
              \frac{4\delta\mu\sqrt{\delta\mu^2-\Delta^2}}{\Delta^2}
             \left(\,-1+2\frac{\delta\mu^2}{\Delta^2}\,\right)\,\right],
 \\[3mm]
 \lambda_{CD} &=&
  -\frac{1}{18\pi^2}\frac{\tilde{\mu}^2}{\Delta^2} \bigg[\, 1 +
    \theta(\delta\mu-\Delta)
    \frac{\delta\mu(3\Delta^2-2\delta\mu^2)}
         {2(\delta\mu^2-\Delta^2)^{\frac{3}{2}}}\,\bigg] .
\end{eqnarray}

In order to calculate the coefficient $T_{DB}$, the cubic term 
in expansion (\ref{V_exp_gII})  is required:
\begin{eqnarray}
   \mathop{\Tr}_{c,f,s}(S_C {\cal M}_{BD})^3 \bigg|_{B^2D} &=&
 -\frac{3}{8}B^2 D \bigg[\,
  \mathop{\Tr}_{f,s} \bigg(\,
  G_\Delta^+(P) \gamma^3 G_b^+(P) \gamma^3 G_\Delta^+(P) \gamma^0
 -G_\Delta^-(P) \gamma^3 G_b^-(P) \gamma^3 G_\Delta^-(P) \gamma^0\,\bigg)
 \nonumber \\ && \qquad
 +\mathop{\tr}_s \bigg(\,
  \gamma^3 \Xi_{12}^-(P) \gamma^0 \Xi_{21}^+(P) \gamma^3
  G_{b}^+(P)\bigg|_{\tau_3=+}
 + \gamma^3 \Xi_{21}^-(P) \gamma^0 \Xi_{12}^+(P) \gamma^3
  G_{b}^+(P)\bigg|_{\tau_3=-}
 \nonumber \\ && \qquad
 - \gamma^3 \Xi_{12}^+(P) \gamma^0 \Xi_{21}^-(P) \gamma^3
  G_{b}^-(P)\bigg|_{\tau_3=+}
 - \gamma^3 \Xi_{21}^+(P) \gamma^0 \Xi_{12}^-(P) \gamma^3
  G_{b}^-(P)\bigg|_{\tau_3=-}
 \,\bigg)\,\bigg] .
 \label{tr_3}
\end{eqnarray}
Then $T_{DB}$ is given by
\begin{equation}
  T_{DB} = \frac{1}{16}\int \frac{d^4 P}{i(2\pi)^4}
  \sum_{e_1=\pm, e_2=\pm, e_3=\pm} \bigg[\,
  {\cal K}_{e_1 e_2 e_3} {\cal T}_{e_1 e_2 e_3}
 +{\cal H}_{e_1 e_2 e_3} {\cal U}_{e_1 e_2 e_3}\,\bigg],
\end{equation}
where
\begin{equation}
  {\cal T}_{e_1 e_2 e_3} \equiv \tr_s \bigg[\,
  \gamma^0 (\gamma^0 \Lambda_p^{e_1})
  \gamma^3 (\gamma^0 \Lambda_p^{e_2})
  \gamma^3 (\gamma^0 \Lambda_p^{e_3})\,\bigg]
\end{equation}
and
\begin{equation}
  {\cal U}_{e_1 e_2 e_3} \equiv \tr_s \bigg[\,
  \gamma^3 (\gamma^0 \Lambda_p^{e_1})
  \gamma^3 (\gamma^5 \Lambda_p^{e_2})
  \gamma^0 (\gamma^5 \Lambda_p^{e_3})\,\bigg] .
\end{equation}
The functions 
${\cal K}_{e_1 e_2 e_3}$ and ${\cal H}_{e_1 e_2 e_3}$
are expressed through the following 
components of the propagators (compare with Eqs. (20)-(24)):
\begin{eqnarray}
  G_{\Delta,\tau}^\pm(P) &=&
  G_{\Delta,+,\tau}^\pm \gamma^0 \Lambda_p^+
 +G_{\Delta,-,\tau}^\pm \gamma^0 \Lambda_p^-, \\
  G_{b,\tau}^\pm(P) &=&
  G_{b,+,\tau}^\pm \gamma^0 \Lambda_p^+
 +G_{b,-,\tau}^\pm \gamma^0 \Lambda_p^-,
\end{eqnarray}
with
\begin{eqnarray}
 G_{\Delta,+,\tau}^\pm(P) &\equiv&
 \frac{(p_0\mp\delta\mu\tau)-E^\pm}
      {(p_0\mp\delta\mu\tau)^2-(E^\pm_\Delta)^2}, \\
 G_{\Delta,-,\tau}^\pm(P) &\equiv&
 \frac{(p_0\mp\delta\mu\tau)+E^\mp}
      {(p_0\mp\delta\mu\tau)^2-(E^\mp_\Delta)^2}, \\
 G_{b,+,\tau}^\pm(P)  &\equiv&
 \frac{1}{(p_0\mp\delta\mu\tau\mp\mu_8)+E^\pm}, \\
 G_{b,-,\tau}^\pm(P)  &\equiv&
 \frac{1}{(p_0\mp\delta\mu\tau\mp\mu_8)-E^\mp} ,
\end{eqnarray}
where $\tau = \pm 1$ are eigenvalues of $\tau_3$, and
\begin{eqnarray}
  \Xi_{12}^\pm (P) &=&
  -i\Delta\left[\,\Xi_{+,\tau=-}^\pm \gamma_5\Lambda_p^+
                 +\Xi_{-,\tau=-}^\pm \gamma_5\Lambda_p^-\,\right], \\
  \Xi_{21}^\pm (P) &=&
  -i\Delta\left[\,\Xi_{+,\tau=+}^\pm \gamma_5\Lambda_p^+
                 +\Xi_{-,\tau=+}^\pm \gamma_5\Lambda_p^-\,\right]
\end{eqnarray}
with
\begin{eqnarray}
  \Xi_{+,\tau}^\pm (P) &\equiv&
  \frac{1}{(p_0 \mp \delta\mu\tau)^2-(E^\pm_\Delta)^2}, \\
  \Xi_{-,\tau}^\pm (P) &\equiv&
  \frac{1}{(p_0 \mp \delta\mu\tau)^2-(E^\mp_\Delta)^2}.
\end{eqnarray}  
The explicit expressions 
of ${\cal K}_{e_1 e_2 e_3}$ and ${\cal H}_{e_1 e_2 e_3}$
are:
\begin{eqnarray}
  {\cal K}_{e_1 e_2 e_3} &=& \sum_{\tau = \pm}\bigg[\,
  G_{\Delta,e_1,\tau}^+ G_{b,e_2,\tau}^+ G_{\Delta,e_3,\tau}^+
 -G_{\Delta,e_1,\tau}^- G_{b,e_2,\tau}^- G_{\Delta,e_3,\tau}^-\,\bigg]
\end{eqnarray}
and
\begin{eqnarray}
  {\cal H}_{e_1 e_2 e_3} &=& -\Delta^2 \bigg[\,
  G_{b,e_1,\tau=+}^+ \Xi_{e_2,\tau=-}^-  \Xi_{e_3,\tau=+}^+
 +G_{b,e_1,\tau=-}^+ \Xi_{e_2,\tau=+}^-  \Xi_{e_3,\tau=-}^+
  \nonumber \\ && \qquad
 -G_{b,e_1,\tau=+}^- \Xi_{e_2,\tau=-}^+  \Xi_{e_3,\tau=+}^-
 -G_{b,e_1,\tau=-}^- \Xi_{e_2,\tau=+}^+  \Xi_{e_3,\tau=-}^-\,\bigg] , \\
&=&
 -\Delta^2 \sum_{\tau = \pm} \bigg[\,
  G_{b,e_1,\tau}^+ \Xi_{-e_2,\tau}^+  \Xi_{e_3,\tau}^+
 -G_{b,e_1,\tau}^- \Xi_{-e_2,\tau}^-  \Xi_{e_3,\tau}^-\,\bigg] ,
\end{eqnarray}
where we used the relations
\begin{equation}
  \Xi_{+,\tau=-}^\pm (P) = \Xi_{-,\tau=+}^\mp (P), \quad
  \Xi_{-,\tau=-}^\pm (P) = \Xi_{+,\tau=+}^\mp (P) .
\end{equation}
We also find
\begin{equation}
  {\cal T}_{+ \pm -} = {\cal T}_{- \pm +} = 0, \quad
  {\cal T}_{+++} = {\cal T}_{---} = 2\frac{(p^3)^2}{p^2}, \quad
  {\cal T}_{+-+} = {\cal T}_{-+-} = 2\left(1-\frac{(p^3)^2}{p^2}\right)
\end{equation}
and
\begin{equation}
  {\cal U}_{\pm ++} = {\cal U}_{\pm --} = 0, \quad
  {\cal U}_{+-+} = {\cal U}_{-+-} = -2\frac{(p^3)^2}{p^2}, \quad
  {\cal U}_{++-} = {\cal U}_{--+} = -2\left(1-\frac{(p^3)^2}{p^2}\right) .
\end{equation}
Thus,
\begin{eqnarray}
  T_{DB} &=& \frac{1}{8}\int \frac{d^4 P}{i(2\pi)^4}
  \bigg[\,\frac{(p^3)^2}{p^2}\bigg(\,
   {\cal K}_{+++} + {\cal K}_{---} - {\cal H}_{+-+} - {\cal H}_{-+-}\,\bigg)
  \nonumber \\ && \qquad \qquad
 +\left(1-\frac{(p^3)^2}{p^2}\right)\bigg(\,
   {\cal K}_{+-+} + {\cal K}_{-+-} - {\cal H}_{++-} - {\cal H}_{--+}\,\bigg)
  \,\bigg], \\[3mm]
&=& \frac{1}{8}\sum_{\tau=\pm}\int \frac{d^4 P}{i(2\pi)^4}\bigg\{
  \frac{(p^3)^2}{p^2}\bigg[\,
  G_{\Delta,-,\tau}^+ G_{b,-,\tau}^+ G_{\Delta,-,\tau}^+
 -G_{\Delta,+,\tau}^- G_{b,+,\tau}^- G_{\Delta,+,\tau}^-
  \nonumber \\ && \hspace*{3.5cm}\quad
 +\Delta^2 \bigg(\,
  G_{b,-,\tau}^+ \Xi_{-,\tau}^+  \Xi_{-,\tau}^+
 -G_{b,+,\tau}^- \Xi_{+,\tau}^-  \Xi_{+,\tau}^-\,\bigg)
  \nonumber \\ && \hspace*{3.5cm}
 +G_{\Delta,+,\tau}^+ G_{b,+,\tau}^+ G_{\Delta,+,\tau}^+
 -G_{\Delta,-,\tau}^- G_{b,-,\tau}^- G_{\Delta,-,\tau}^-
  \nonumber \\ && \hspace*{3.5cm}\quad
 +\Delta^2 \bigg(\,
  G_{b,+,\tau}^+ \Xi_{+,\tau}^+  \Xi_{+,\tau}^+
 -G_{b,-,\tau}^- \Xi_{-,\tau}^-  \Xi_{-,\tau}^-\,\bigg)\,\bigg]
  \nonumber \\ && \hspace*{1.5cm}
 +\left(1-\frac{(p^3)^2}{p^2}\right)\bigg[\,
  G_{\Delta,-,\tau}^+ G_{b,+,\tau}^+ G_{\Delta,-,\tau}^+
 -G_{\Delta,+,\tau}^- G_{b,-,\tau}^- G_{\Delta,+,\tau}^-
  \nonumber \\ && \hspace*{3.5cm}\quad
 +\Delta^2 \bigg(\,
  G_{b,+,\tau}^+ \Xi_{-,\tau}^+  \Xi_{-,\tau}^+
 -G_{b,-,\tau}^- \Xi_{+,\tau}^-  \Xi_{+,\tau}^-\,\bigg)
  \nonumber \\ && \hspace*{3.5cm}
 +G_{\Delta,+,\tau}^+ G_{b,-,\tau}^+ G_{\Delta,+,\tau}^+
 -G_{\Delta,-,\tau}^- G_{b,+,\tau}^- G_{\Delta,-,\tau}^-
  \nonumber \\ && \hspace*{3.5cm}\quad
 +\Delta^2 \bigg(\,
  G_{b,-,\tau}^+ \Xi_{+,\tau}^+  \Xi_{+,\tau}^+
 -G_{b,+,\tau}^- \Xi_{-,\tau}^-  \Xi_{-,\tau}^-\,\bigg)\,\bigg] .
\end{eqnarray}
After integrating over $p^0$, we find
\begin{equation}
  T_{DB} = \frac{1}{24\pi^2}\int_0^\Lambda dp\,p^2\,
  \bigg[\,I_1 + 2 I_2\,\bigg]
\end{equation}
with
\begin{eqnarray}
  I_1 &\equiv&
  -\frac{E_\Delta^- - E^-}{E_\Delta^-(E_\Delta^- + E^- + \mu_8)^2}
  +\frac{E_\Delta^+ - E^+}{E_\Delta^+(E_\Delta^+ + E^+ - \mu_8)^2}
   \nonumber \\[3mm] &&
  +\frac{1}{2}\Big(\,\theta(-E^- + \delta\mu - \mu_8)
                    +\theta(-E^- - \delta\mu - \mu_8)\,\Big)
   \left(\,\frac{E_\Delta^- - E^-}{E_\Delta^-(E_\Delta^- + E^- + \mu_8)^2}
          +\frac{E_\Delta^- + E^-}{E_\Delta^-(E_\Delta^- - E^- - \mu_8)^2}
   \,\right)
   \nonumber \\[3mm] &&
  +\frac{1}{2}\theta(-E_\Delta^- + \delta\mu)
   \left(\,\frac{E_\Delta^- - E^-}{E_\Delta^-(E_\Delta^- + E^- + \mu_8)^2}
          -\frac{E_\Delta^- + E^-}{E_\Delta^-(E_\Delta^- - E^- - \mu_8)^2}
   \,\right)
\end{eqnarray}
and
\begin{eqnarray}
  I_2 &\equiv&
   \frac{E_\Delta^- + E^-}{E_\Delta^-(E_\Delta^- + E^+ - \mu_8)^2}
  -\frac{E_\Delta^+ + E^+}{E_\Delta^+(E_\Delta^+ + E^- + \mu_8)^2}
   \nonumber \\[3mm] &&
  +\frac{1}{2}\Big(\,\theta(-E^- + \delta\mu - \mu_8)
                    +\theta(-E^- - \delta\mu - \mu_8)\,\Big)
   \left(\,\frac{E_\Delta^+ + E^+}{E_\Delta^+(E_\Delta^+ + E^- + \mu_8)^2}
          +\frac{E_\Delta^+ - E^+}{E_\Delta^+(E_\Delta^+ - E^- - \mu_8)^2}
   \,\right)
   \nonumber \\[3mm] &&
  +\frac{1}{2}\theta(-E_\Delta^- + \delta\mu)
   \left(\,\frac{E_\Delta^- - E^-}{E_\Delta^-(E_\Delta^- - E^+ + \mu_8)^2}
          -\frac{E_\Delta^- + E^-}{E_\Delta^-(E_\Delta^- + E^+ - \mu_8)^2}
   \,\right).
\end{eqnarray}
The integrand $I_2$ contains only the particle-antiparticle
contribution and hence should be negligible.
We expand $I_{1,2}$ in the series with respect to $\mu_8$ and 
obtain
\begin{eqnarray}
  T_{DB} &=& \frac{\tilde{\mu}}{48\pi^2}
  \bigg[\,-1 + 4\frac{\delta\mu^2}{\Delta^2} + 8\frac{\delta\mu^4}{\Delta^4}
  -\theta(\delta\mu-\Delta)
    \frac{8\delta\mu (\delta\mu^2-\Delta^2)^{\frac{3}{2}}}{\Delta^4}\,\bigg]
 \nonumber \\ &&
 +\frac{\mu_8}{24\pi^2}\frac{\tilde{\mu}^2}{\Delta^2}
  \bigg[\,-1 + 8\frac{\delta\mu^4}{\Delta^4}
  -\theta(\delta\mu-\Delta)
    \frac{4\delta\mu(2\delta\mu^2-\Delta^2)\sqrt{\delta\mu^2-\Delta^2}}
         {\Delta^4}\,\bigg] .
\end{eqnarray}
As was expected, the contribution of $I_2$ is suppressed indeed.


\begin{thebibliography}{99}
\bibitem{footnote1} Here ``sufficiently large values of
$\mu_Y$'' means the following: When $m^2 > 0$, $\mu_Y$ should be larger
than the critical value $\mu_{cr} = m$, and for $m^2 < 0$,
$\mu_Y$ should be larger
than $\mu_{cr} = g|m|/2\sqrt{\lambda}$ (the
critical value $g|m|/2\sqrt{\lambda}$ coincides with the mass
of $W$ boson in the vacuum theory with $\mu_Y =0$ and $m^2 <0$).
\bibitem{footnote}
 It is a partly (and not strictly) unitary gauge because
it eliminates only those $SU(3)_c$ gauge transformations 
which are assigned to the coset space $SU(3)_c/SU(2)_c$.
\end{thebibliography}
\end{document}